\documentclass[10pt,twocolumn,letterpaper]{article}

\usepackage{iccv}              % To produce the CAMERA-READY version
\usepackage{multirow}
\usepackage{booktabs}
\usepackage{array}
\usepackage{calc}
\usepackage{graphicx}

\newcommand{\best}[1]{\textcolor{red}{#1}}
\usepackage{amsmath}

\usepackage{makecell}

\definecolor{iccvblue}{rgb}{0.21,0.49,0.74}
\usepackage[pagebackref,breaklinks,colorlinks,allcolors=iccvblue]{hyperref}

\def\paperID{3377} % *** Enter the Paper ID here
\def\confName{ICCV}
\def\confYear{2025}

\title{CDI: Blind Image Restoration Fidelity Evaluation based on Consistency with Degraded Image}

\author{Xiaojun Tang \and Jingru Wang \and Guangwei Huang \and Guannan Chen \and Rui Zheng \and Lian Huai \and Yuyu Liu \and Xingqun Jiang
}
\begin{document}
\maketitle
\begin{abstract}
Recent advances in Blind Image Restoration (BIR) methods, based on generative adversarial networks and diffusion models, have significantly improved visual quality. However, they pose significant challenges for Image Quality Assessment (IQA), as conventional Full-Reference IQA methods often assign low scores to images with high perceptual quality. In this paper, we reassess the issues of Solution Non-Uniqueness and Degradation Indeterminacy in BIR, and propose a novel IQA system based on Consistency with Degraded Image (CDI). Rather than comparing a restored image with a reference image, CDI compares the restored image with the degraded image, enabling accurate fidelity assessment. Specifically, we propose a wavelet domain Reference Guided CDI algorithm, which effectively quantifies fidelity across various types of degraded images without requiring prior knowledge of the degradation. The supported degradation types include down sampling, blur, noise, JPEG artifacts, complex combined degradations, etc. In addition, we propose a Reference Agnostic CDI, allowing fidelity evaluation without reference images. Finally, we construct a new Degraded Images Switch Display Comparison Dataset (DISDCD) for subjective fidelity evaluation. Experiments verify that CDI is markedly superior to conventional Full Reference IQA methods in assessing fidelity. The source code and the dataset will be publicly released in the near future.
\end{abstract}    
\section{Introduction}
\label{sec:intro}

Traditional image restoration algorithms, including image denoising \cite{DNCNN, DRUNet}, deblurring \cite{BSSTNet, GShift-Net} and super-resolution \cite{RCAN, SAN, HAN} \textit{et al.}, are designed and trained to restore images with specifically known degradation. However, these methods typically present limited generalization ability and struggle to restore images degraded by complex or unknown distortions. In recent years, Blind Image Restoration (BIR) \cite{RealESRGAN, BSRGAN, DiffBIR, StableSR, SeeSR} has emerged as a promising research direction, aiming to reconstruct images degraded by unknown real-world distortions. Notably, BIR algorithms possess the capability to generate fine details, and can produce restoration results with rich textures even for severely degradations. Due to the uncertainty of whether the generated details are correct, it is necessary to find an accurate assessment of the fidelity.

\begin{figure}[t]
\begin{subfigure}{.85\linewidth}
    \centering
    \includegraphics[width=85mm]{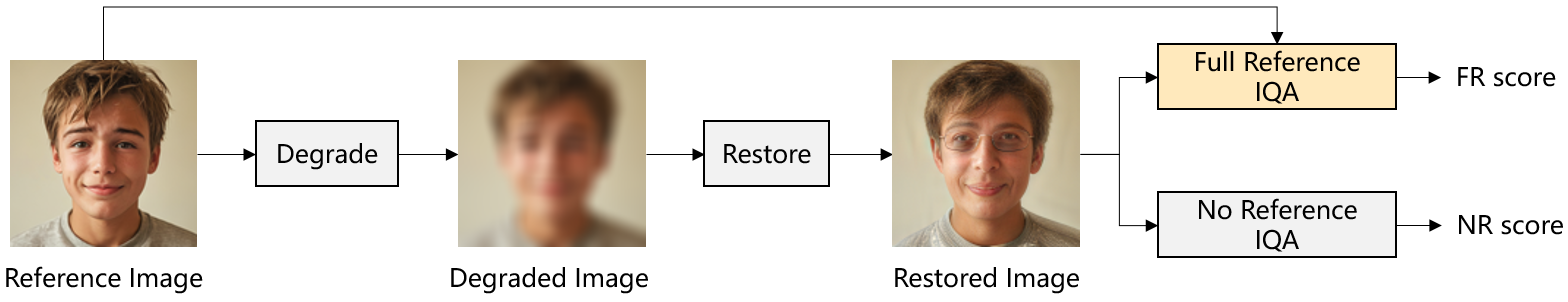}
    \caption{Classic IQA}
    \label{fig:iqa_a}
\end{subfigure}
\qquad
\begin{subfigure}{.85\linewidth}
    \centering
    \includegraphics[width=85mm]{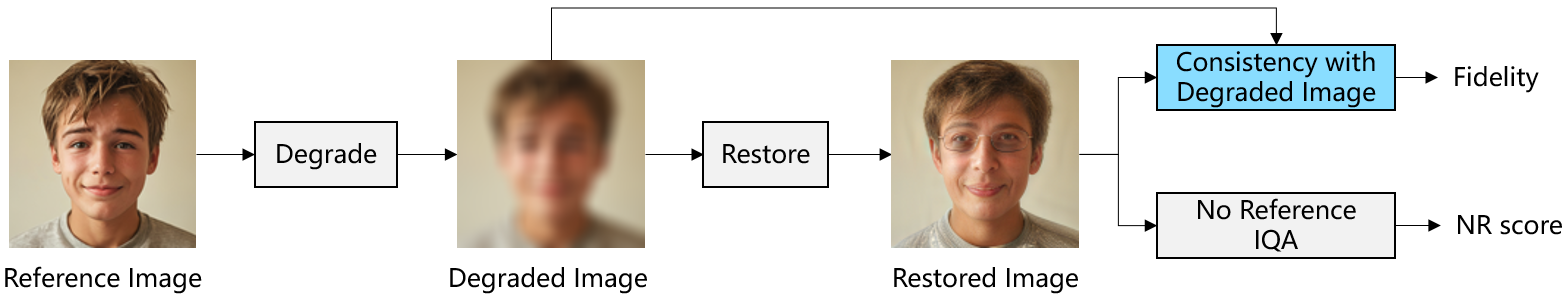}
    \caption{Proposed IQA}
    \label{fig:iqa_b}
\end{subfigure}
  \caption{(a) Classic IQA includes Full Reference IQA and No Reference IQA. (b) The proposed IQA evaluates fidelity by Consistency with Degraded Image (CDI). (To avoid infringing on facial portrait rights, we choose an image generated by Stable Diffusion as the reference image.)}
  \label{fig:iqa}
\end{figure}

\subsection{Problems of Fidelity Evaluation}
\label{subsec:problems of fidelity evaluation}

\begin{figure}[t]
  \centering
  \includegraphics[width=85mm]{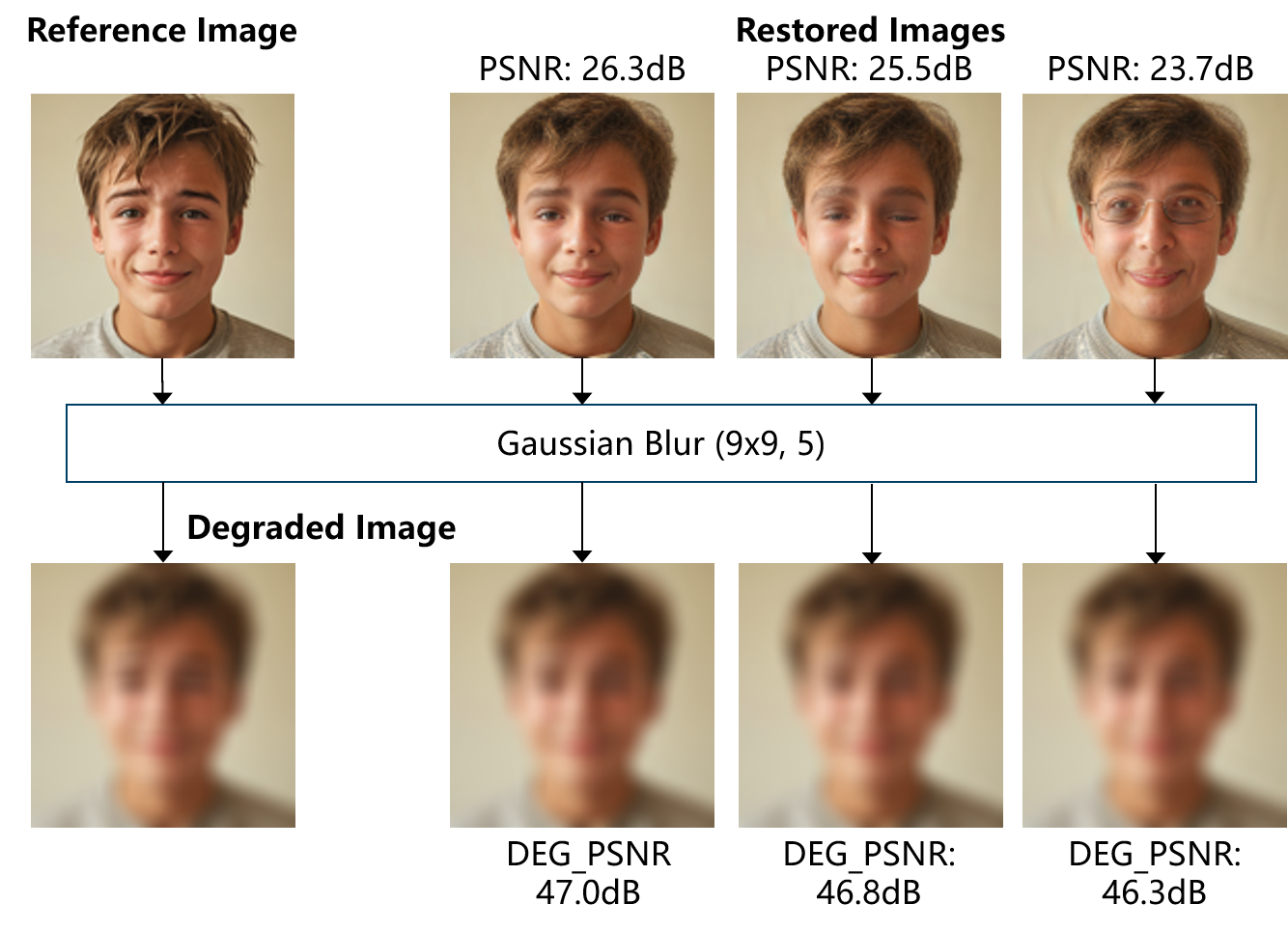}
  \caption{BIR resolution is non-unique. It is impossible to determine which one is the ground truth reference image given only the degraded image. Note: $DEG\_PSNR$ represents the PSNR of two degraded images. }
  \label{fig:resolution_non-unique}
\end{figure}

\noindent \textbf{BIR Solution Non-Uniqueness.} 
Image degradation often leads to irreversible loss of image information. As its inverse process, image restoration is an ill-posed problem. All restored images that can be degraded to the low quality image should be considered as correct solutions. Fig.\ref{fig:resolution_non-unique} shows that given an image blurred by a Gaussian filter, three visually distinct images are generated using existing BIR algorithms $($with $DEG\_PSNR$ maximization gradient adjustment$)$. The first image has different hair textures, the second has closed eyes, and the third shows an individual wearing glasses. Evidently, all three images have low PSNR scores. However, once the same Gaussian blur is applied, they appear near-identical to the degraded image $(DEG\_PSNR>46dB)$, indicating that they are all valid solutions. In fact, the reference image is merely one solution within the aforementioned ill-posed solution space. It is inappropriate to simply use the consistency with reference image as an evaluation metric, especially when only the degraded image is available.

\begin{figure}[t]
  \centering
  \includegraphics[width=85mm]{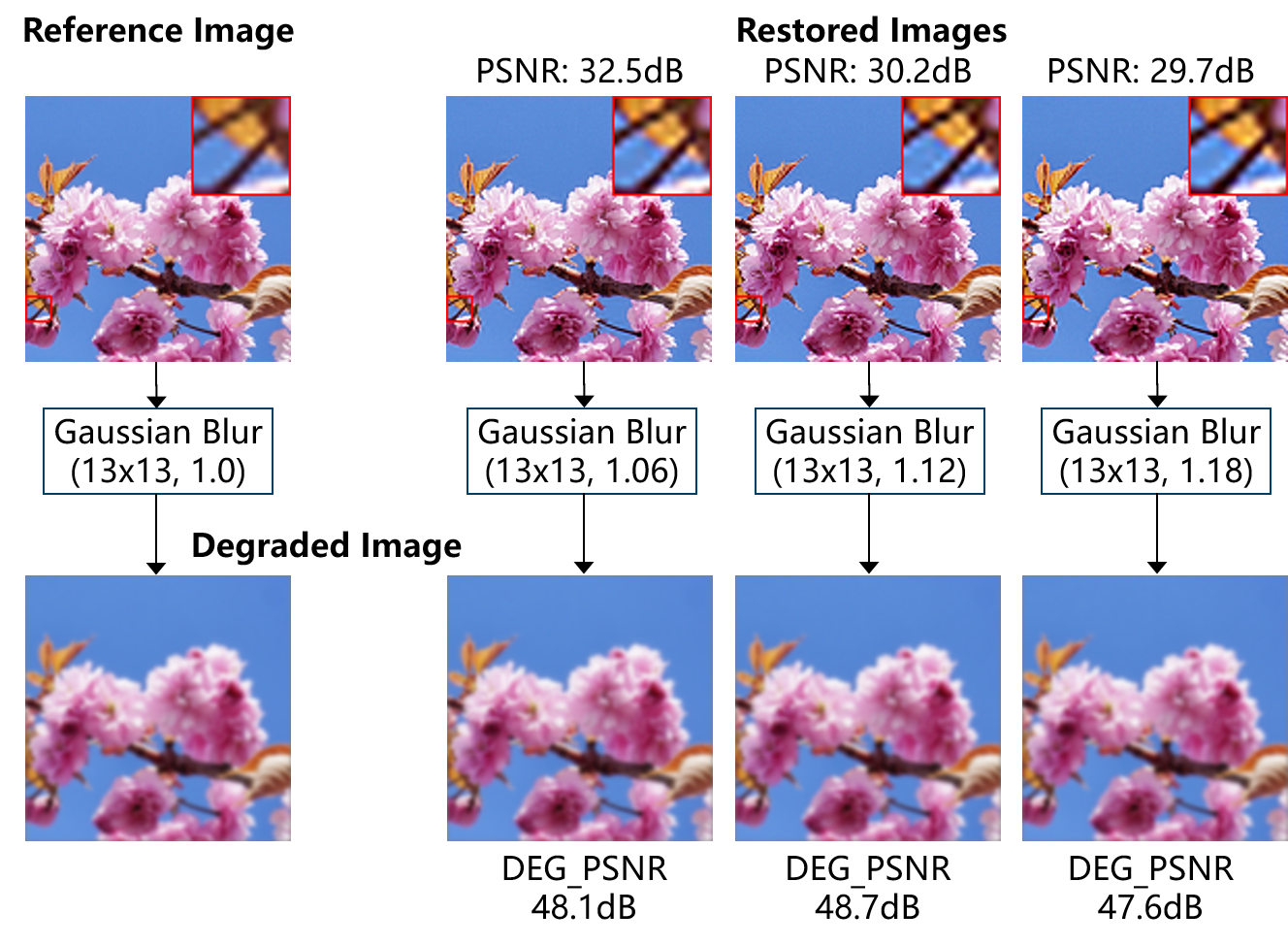}
  \caption{Given the indeterminate nature of BIR degradation, it's impossible to accurately determine the degradation parameters solely through an examination of the degraded image.}
  \label{fig:degradation_Indeterminacy}
\end{figure}

\noindent \textbf{BIR Degradation Indeterminacy.} 
For degradation-known image restoration, an intuitive idea is to apply the same degradation process to the restored image and then compare it with the degraded image. However, accurate degradation parameters are inaccessible in BIR. And once the degradation parameters in the inverse reconstruction process are altered, the corresponding solution space will change accordingly. As shown in Fig.\ref{fig:degradation_Indeterminacy}, given a blurred image degraded by the Gaussian kernel $\sigma\!=\!1.0$, we generate three different images by varying the value $\sigma$ in the inverse reconstruction process. After applying a Gaussian blur with different $\sigma$ values, these three images closely resemble the degraded image. In other words, they are all valid solutions. Therefore, the solution space of the BIR should not be constrained by strict degradation parameters.

\begin{figure}[h]
\begin{subfigure}{.85\linewidth}
    \centering
    \includegraphics[width=85mm]{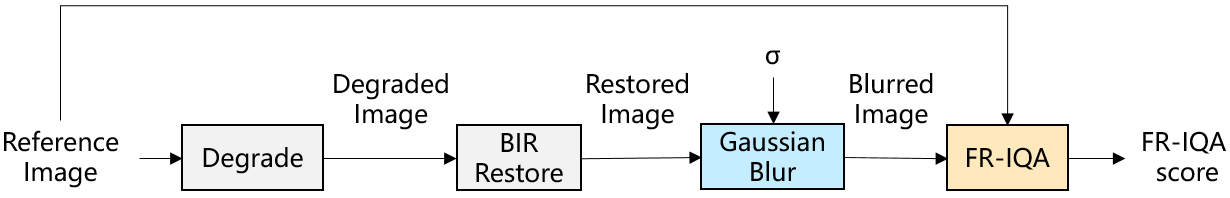}
    \caption{Experimental System}
    \label{fig:bir_blur_test_system}
\end{subfigure}

\begin{subfigure}{.85\linewidth}
    \centering
    \includegraphics[width=55mm]{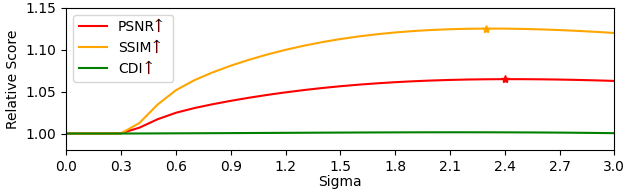}
    \caption{Test Results of PNSR, SSIM \cite{SSIM} and CDI}
    \label{fig:blur_test_2}
\end{subfigure}

\begin{subfigure}{.85\linewidth}
    \centering
    \includegraphics[width=55mm]{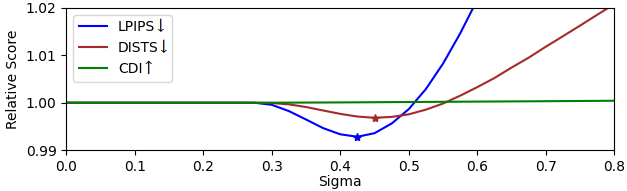}
    \caption{Test Results of LPIPS \cite{LPIPS}, DISTS \cite{DISTS} and CDI}
    \label{fig:blur_test_3}
\end{subfigure}

  \caption{(a) BIR Gussian blur test system. (b, c) Appropriate Gaussian blur can improve PSNR, SSIM, LPIPS and DISTS scores. The $RGCDI\_PSNR$ proposed in this paper is almost unaffected. Note: Real-ESRGAN degradation \cite{RealESRGAN} are adopted on DIV2K \cite{DIV2K} validation dataset to get degraded images.}
  \label{fig:bir_blur_test}
\end{figure}

\noindent \textbf{FR-IQA tends to give better scores to blurred images.} Although Full Reference IQA (FR-IQA) algorithms \cite{SSIM, LPIPS, DISTS} are currently the standard methods for evaluating image fidelity, numerous experiments conducted on the DIV2K \cite{DIV2K} validation dataset have demonstrated that the scores of BIR algorithms \cite{RealESRGAN, BSRGAN, DiffBIR, StableSR, SeeSR}  are inferior. While higher quality restored images tend to get worse FR-IQA scores, blurred images, on the contrary, obtain better scores. Fig.\ref{fig:bir_blur_test_system} shows that we perform Gaussian blur (standard deviation $\sigma$) on the BIR restored image and then calculate the FR-IQA scores of the blurred image. The curves of different FR-IQA scores as a function of $\sigma$ are plotted in Fig.\ref{fig:blur_test_2} and Fig.\ref{fig:blur_test_3}, indicating that adding appropriate blur helps improve the FR-IQA scores. For each FR-IQA algorithm, the scores are normalized using its result on non-blurred images.

The reasons for this ill phenomenon are as follows: severe degradation results in the loss of a large amount of high-frequency details. The high-frequency details generated by the BIR algorithms often differ from those of the reference image, leading to a substantial decrease in FR-IQA scores. However, these details may also belong to the non-unique solution space. Although indiscriminate blurring of these high-frequency components can improve FR-IQA scores, it also compromises image quality, which is undesirable. Consequently, existing FR-IQA algorithms are not suitable for the assessment of BIR image fidelity.

In contrast, the proposed Reference Guided Consistency with Degraded Image ($RGCDI\_PSNR$) keeps largely unaffected by Gaussian blur and provides a more accurate evaluation of fidelity (Fig.\ref{fig:blur_test_2} and Fig.\ref{fig:blur_test_3}).

\subsection{Consistency with Degraded Image}

Fig.\ref{fig:iqa_a} illustrates the structure of the classic Image Quality Assessment (IQA) \cite{SSIM, LPIPS, DISTS, NIQE, MANIQA, BRISQUE, PI}. In this system, Full Reference IQA (FR-IQA) is designed to estimate the pixel-level or perceptual similarity between the restored images and the reference images. In contrast, No Reference IQA (NR-IQA) evaluates the quality of restored images based on prior knowledge, without the need for reference images.

As discussed in Sec.\ref{subsec:problems of fidelity evaluation}, FR-IQA is not a rational BIR fidelity metrics. Therefore, we proposed the CDI-IQA system, as illustrated in Fig.\ref{fig:iqa_b}, which evaluates CDI by comparing restored images with degraded images. From the perspective of information theory, image degradation leads to the loss of high-frequency information, leaving degraded images with only low-frequency components. Once high-frequency information is lost, it cannot be recovered. Image restoration can only generate visually plausible high-frequency information based on probability. Therefore, CDI should disregard high-frequency information and focus on assessing the consistency of low-frequency information.

It should also be noted that CDI only evaluates the fidelity of low-frequency information and does not check the high-frequency signals generated by image restoration. The prior image quality of high-frequency signals in restored images can be evaluated using NR-IQA. CDI fidelity and NR-IQA constitute two orthogonal and complementary metrics for BIR Image Quality Assessment.

\subsection{The Main Work of This Paper}

In this paper, we focus on CDI based fidelity evaluation. In Sec.\ref{sec:related}, we review the existing related research work. In Sec.\ref{sec:friqa_on_degraded}, we analyze the problems of calculating FR-IQA on degraded images as a CDI metrics, particularly the inability to handle noise. In Sec.\ref{sec:subjective_modeling}, we tackle the challenge of fidelity evaluation under noise degradation by fidelity subjective evaluation modeling. Using an information-theoretic model, we separately quantify and analyze the impact of signal attenuation and noise on the amount of transmitted information. In Sec.\ref{sec:rgcdi}, we propose a wavelet-based Reference Guided CDI (RGCDI) algorithm using subjective evaluation modeling, and realize accurate CDI calculation with reference images. In Sec.\ref{sec:racdi}, we train an Attenuation Net and achieve the Reference Agnostic CDI (RACDI) calculation. In Sec.\ref{sec:dataset}, we build a new Degraded Images Switch Display Comparison Dataset (DISDCD) for subjective evaluation of BIR fidelity. To the best of our knowledge, this is the first IQA dataset of fidelity evaluation based on degraded images. Finally, the experimental analysis is presented in Sec.\ref{sec:experiments}, followed by conclusions in Sec.\ref{sec:conclusion}.

The contribution of this work is summarized as follows:

\begin{itemize}
[itemindent = 0pt, labelindent = \parindent, labelwidth = 2em, labelsep = 5pt, leftmargin = *]
   \item We systematically analyze the problems of BIR fidelity evaluation and propose a new fidelity evaluation architecture based on CDI.

   \item By Human Vision System (HVS) information modeling, we propose replacing noise degradation with information-equivalent signal attenuation to improve fidelity calculation.
   
   \item A wavelet-based Reference Guided CDI (RGCDI) algorithm is proposed, which can evaluate CDI scores for various types of degradations without knowing the degradation information.
   
   \item We train a general Attenuation Net, achieving the Reference Agnostic CDI (RACDI) calculation.

   \item We built the first IQA dataset for fidelity evaluation based on degraded images, named DISDCD. Experiments verify that the proposed RGCDI and RACDI are significantly superior to other IQA algorithms in terms of fidelity evaluation.
\end{itemize}
\section{Related work}
\label{sec:related}

\noindent \textbf{Full-Reference Image Quality Assessment $($FR-IQA$)$} methods, such as PSNR, SSIM \cite{SSIM}, LPIPS \cite{LPIPS}, DISTS \cite{DISTS} \textit{et al.}, measure the similarity between two images. The early FR-IQA methods \cite{SSIM} mainly calculate pixel level similarity, while the recent FR-IQA methods \cite{LPIPS, DISTS} focus on perceptual similarity. All of them do not consider degraded images and are inappropriate in severe degradation scenarios.

\noindent \textbf{No-Reference Image Quality Assessment $($NR-IQA$)$} methods, for example, NIQE \cite{NIQE}, MANIQA \cite{MANIQA}, BRISQUE \cite{BRISQUE}, and PI \cite{PI} \textit{et al.}, are proposed to assess perceptual image quality without reference image. They do not evaluate fidelity.

\noindent \textbf{Reduced-Reference Image Quality Assessment $($RR-IQA$)$} methods, such as RRED \cite{RRED}, OSV \cite{OSVP}, evaluate image quality by comparing the partial information of the reference image. Similarly to FR-IQA methods, RR-IQA methods also do not take into account degraded images.

\noindent \textbf{IQA datasets}, such as LIVE \cite{LIVE}, CSIQ \cite{CSIQ}, TID2008 \cite{TID2008}, TID2013 \cite{TID2013}, PieAPP \cite{PieAPP}, and BAPPS \cite{BAPPS}, QADS \cite{QADS}, PIPAL \cite{PIPAL}, are often used to test the performance of IQA methods. They provide subjective scores for various distorted images. Since all existing datasets are based on reference images instead of degraded images, they are not suitable to assess the performance of fidelity evaluation (Sec.\ref{subsec:problems of fidelity evaluation}). A new dataset with fidelity based on degraded images needs to be constructed.

% \noindent \textbf{Plug-and-play Image Restoration} includes a series of new promising image restoration methods \cite{K-SVD, BM3D, DPIR, DIFFPIR, DDNM}. Generally, they adopt a two step image degradation modeling and solve image restoration problems by calculating a Maximum A Posterior (MAP). Then, the image restoration target is split into two items, minimizing the error of degraded images and maximizing the prior probability of image. The former can be regarded as a good fidelity quantitative evaluation index. However, existing Plug-and-play methods depend on known degradation operations, cannot be used in Blind Image Restoration. In this paper, the proposed Consistency with Degraded Image methods do not rely on degradation information and can be used for BIR fidelity evaluation.
%-------------------------------------------------------------------------
\section{FR-IQA on Degraded Image}
\label{sec:friqa_on_degraded}

\begin{figure}[h]
  \centering
  \includegraphics[width=55mm]{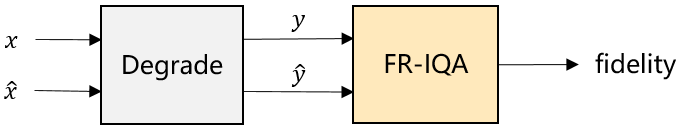}
  \caption{FR-IQA on Degraded Image}
  \label{fig:fr_on_degradation}
\end{figure}

The most intuitive method of calculating CDI is to apply the established FR-IQA algorithms to the degraded images. As Fig.\ref{fig:fr_on_degradation} illustrated, we apply the same degradation to the restored image under evaluation and then calculate the consistency score using FR-IQA. While this scheme works well for degradations with low noise intensity (e.g. blur and down sampling), it exhibits significant limitations in high noise intensity degradation scenarios (e.g. JPEG and Gaussian noise). The quantization noise in JPEG compression is uncontrollable, even minor changes to the reference image can result in significant variations, thereby affecting the evaluation of fidelity. Gaussian noise can be controlled by fixing random seeds, but the performance of FR-IQA on Gaussian noise is also poor.

FR-IQA methods based on human perception (e.g., LPIPS, DISTS) fail to achieve accurate fidelity scores because they are not trained on noisy reference images. Furthermore, obtaining reference images is not feasible in real-world image degradation scenarios, meaning that the FR-IQA on Degraded Image scheme cannot be extended to real-world BIR.

Experimental results in Fig.\ref{fig:2AFC} show that FR-IQA on Degraded Image perform poorly when applied to Gaussian noise, JPEG compression and combined degradation patterns. Therefore, it is essential to investigate the impact of noise and optimize the fidelity evaluation for images restored from noisy degradation.

%-------------------------------------------------------------------------
\section{Fidelity Subjective Evaluation Modeling}
\label{sec:subjective_modeling}

To analyze the impact of noise on fidelity evaluation, we first use a two-step image degradation model to separate image degradation into image attenuation and noise (Sec.\ref{subsec:two_step_degrade}). We then examine the effects of attenuation and noise on subjective perception of fidelity individually (Sec.\ref{subsec:hvs_model}).

\subsection{Two-Step Image Degradation Modeling}
\label{subsec:two_step_degrade}

\begin{figure}[h]
  \centering
  \includegraphics[width=65mm]{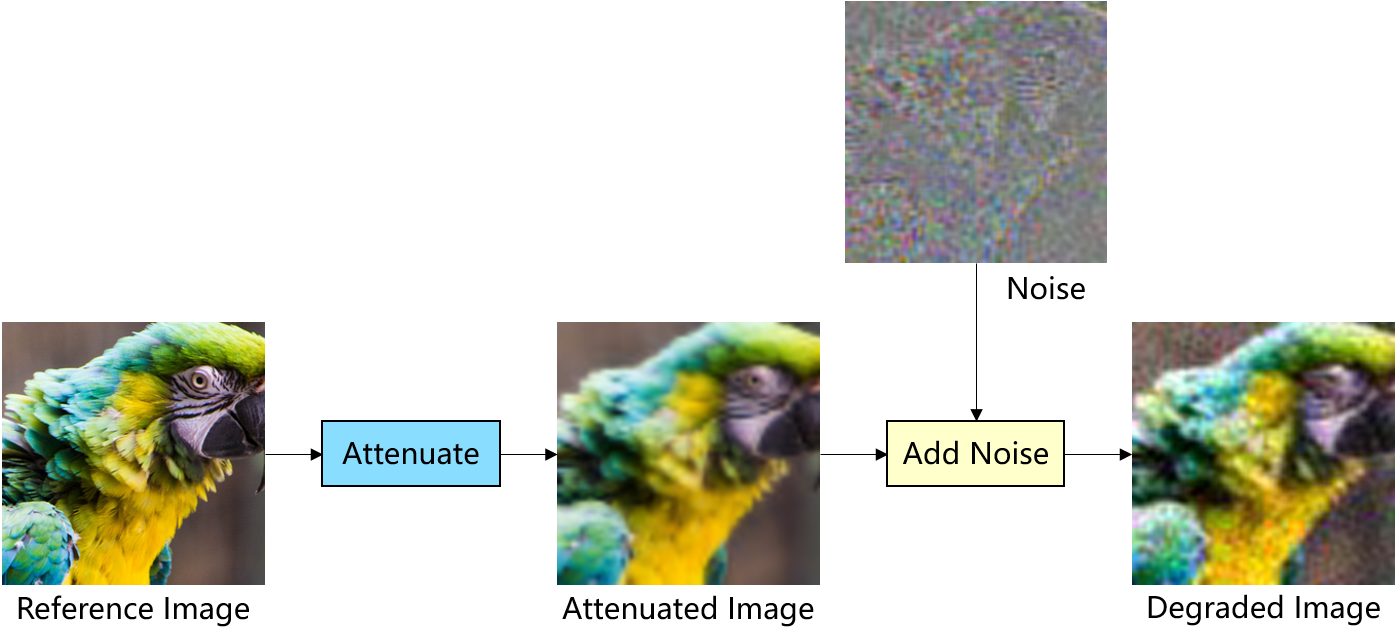}
  \caption{Two-Step Image Degradation Modeling}
  \label{fig:DegradationModeling}
\end{figure}

Image degradation typically involves the loss of high-frequency details and the addition of noise, which can be modeled using a general two-step degradation process \cite{DPIR, DIFFPIR, DDNM} ${I}_{y}=\Gamma \left ( {I}_{x} \right ) +{I}_{n}$, where $\Gamma$ is a data attenuation operation and ${I}_{n}$ is assumed to be additive noise, as Fig.\ref{fig:DegradationModeling} illustrated.

Similarly, in the wavelet domain, image degradation can also generally be modeled using the two-step degradation process \cite{VIF} as shown in Eq.\ref{formula:DM_1}. In this model, ${\mu}_{A}$ represents the attenuation coefficient and ${n}_{D}$ is assumed to be an additive Gaussian noise with a standard deviation of ${\sigma}_{D}$. The image wavelet coefficient $x$ can be modeled using Gaussian Scale Mixture (GSM) \cite{GSM}, as expressed in Eq.\ref{formula:DM_2}, where $s$ is a Random Field (RF) of positive scalars and $u$ is a Gaussian vector RF with mean zero and covariance ${C}_{u}$. This model effectively simulates the attenuation and noise of high-frequency signals across different frequencies and adequately captures the effects of distortions in the real world in terms of perceptual annoyance \cite{VIF}.

\begin{equation}\label{formula:DM_1}
    y = {\mu}_{A} x + {n}_{D}
\end{equation}

\begin{equation}\label{formula:DM_2}
    x = s \cdot u
\end{equation}

According to Eq.\ref{formula:DM_1}, we aim to decompose $y$ into ${\mu}_{A} x$ and ${n}_{D}$. Given reference $x$, the noise separation is straightforward, as $COV \left ({n}_{D}, x\right ) = 0 $ (that is, $x$ and ${n}_{D}$ are independent in probability \cite{VIF}, see Eq.\ref{formula:WNS}). The standard deviation $\sigma_{n}$ can also be estimated \cite{VIF}.

\begin{equation}\label{formula:WNS}
\begin{aligned}
    &{\mu}_{A} = COV \left( y, x \right) / COV \left ( x,x \right )  \\
    &{n}_{D} = y - {\mu}_{A} x  \\
    &\sigma_{D}^{2} = COV \left (y, y \right ) - {\mu}_{A} \cdot COV \left (y, x \right )
\end{aligned}
\end{equation}

\begin{figure}[h]
  \centering
  \includegraphics[width=85mm]{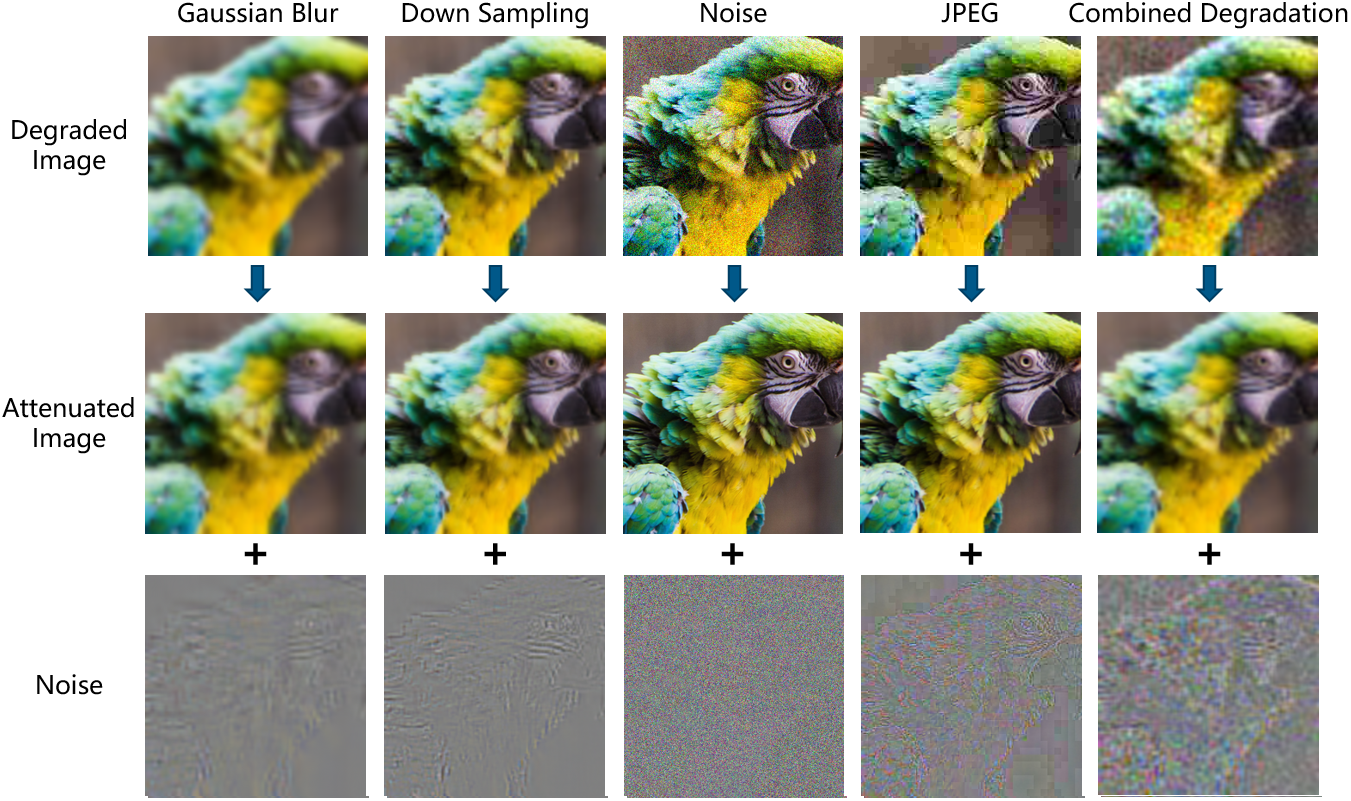}
  \caption{Example sample images of Wavelet Noise Splitting}
  \label{fig:WNS}
\end{figure}

Fig.\ref{fig:WNS} presents split images of various degradations. Gaussian blur and down sampling exhibit significant high-frequency attenuation with little noise. In contrast, Gaussian noise is mainly composed of additional noise with little attenuation. JPEG and combined degradation \cite{RealESRGAN} involve both attenuation and noise.

\subsection{Human Vision System Information Modeling}
\label{subsec:hvs_model}
\begin{figure}[h]
\begin{subfigure}{.85\linewidth}
    \centering
    \includegraphics[width=65mm]{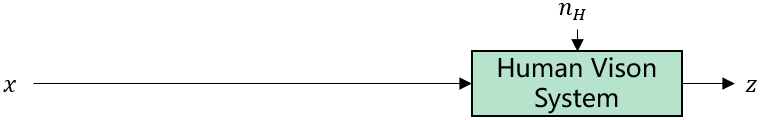}
    \caption{HVS without degradation}
    \label{fig:hvs_model_1}
\end{subfigure}
\qquad

\vspace{0.5em}
\begin{subfigure}{.85\linewidth}
    \centering
    \includegraphics[width=65mm]{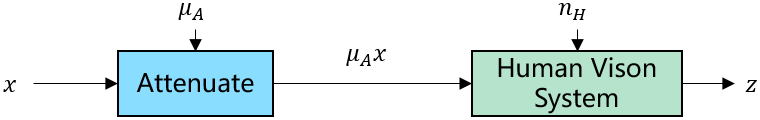}
    \caption{HVS with attenuation}
    \label{fig:hvs_model_2}
\end{subfigure}
\qquad

\vspace{0.5em}
\begin{subfigure}{.85\linewidth}
    \centering
    \includegraphics[width=65mm]{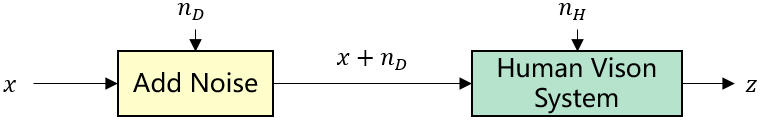}
    \caption{HVS with noise}
    \label{fig:hvs_model_3}
\end{subfigure}
\qquad

\vspace{0.5em}
\begin{subfigure}{.85\linewidth}
    \centering
    \includegraphics[width=65mm]{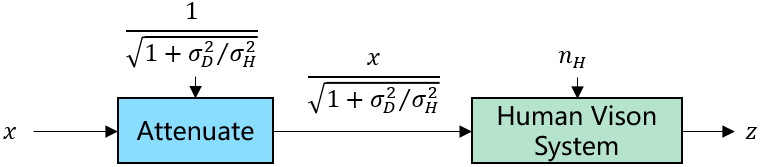}
    \caption{HVS with noise equivalent attenuation}
    \label{fig:hvs_model_4}
\end{subfigure}

  \caption{Human Vision System (HVS) Information Modeling.}
  \label{fig:CDI}
\end{figure}

First, we assume that there is no image degradation, as illustrated in Fig.\ref{fig:hvs_model_1}. The Human Vision System (HVS) can be modeled as a noisy information transmission system \cite{VIF}. Its output is calculated as follows: $z = x + {n}_{H}$, where $x$ represents the wavelet coefficient of the original image, and can be modeled as a Gaussian variable with covariance $COV \left ( x,x \right )$. ${n}_{H}$ represents the additive noise introduced by HVS, and can be modeled as a Gaussian variable with covariance ${\sigma}_{H}^{2}$, where ${\sigma}_{H}$ denotes the intensity of the noise. The amount of information transmitted in HSV is represented by mutual information $I \left ( x,z \right ) = H \left ( x \right ) - H \left ( x | z \right )$, which can be calculated as shown in Eq.\ref{formula:mutual_information_1}. It is evident that a smaller ${\sigma}_{H}$ indicates a higher signal-to-noise ratio, which in turn leads to increased information transmission in the HVS.

Second, when signal attenuation is introduced (Fig.\ref{fig:hvs_model_2}), the mutual information is calculated as shown in Eq.\ref{formula:mutual_information_2}. Attenuation leads to a decrease in signal-to-noise ratio, thereby reducing the amount of information received.

Third, when noise is introduced (Fig.\ref{fig:hvs_model_3}), the mutual information is calculated as shown in Eq.\ref{formula:mutual_information_3}. Noise also leads to a decrease in signal-to-noise ratio, thereby reducing the amount of information received.

Finally, since both signal attenuation and noise contribute to information loss, we can convert noise into equivalent signal attenuation that results in the loss of the same amount of information. HVS with noise-equivalent attenuation is illustrated in Fig.\ref{fig:hvs_model_4}. The mutual information is calculated as shown in Eq.\ref{formula:mutual_information_4}, and it can be proven that the values calculated from Eq.\ref{formula:mutual_information_3} and Eq.\ref{formula:mutual_information_4} are identical.

\begin{subequations}
\begin{align}
    I_1(x,z) = log \frac{ | COV ( x, x ) + {\sigma}_{H}^{2} |}{ | {\sigma}_{H}^{2} |} \label{formula:mutual_information_1}
\end{align}

\begin{align}
    I_2(x,z) = log \frac{ | COV ( {\mu}_{A} x, {\mu}_{A} x ) + {\sigma}_{H}^{2}|}{ | {\sigma}_{H}^{2}|} \label{formula:mutual_information_2}
\end{align}

\begin{align}
    I_3(x,z) = log \frac{ | COV ( x, x ) + {\sigma}_{D}^{2} + {\sigma}_{H}^{2}|}{ | {\sigma}_{D}^{2} + {\sigma}_{H}^{2}|} \label{formula:mutual_information_3}
\end{align}

\begin{align}
    I_4(x,z) = log \frac{ | COV ( \frac{x}{\sqrt{1 + {\sigma}_{D}^{2}/{\sigma}_{H}^{2}}}, \frac{x}{\sqrt{1 + {\sigma}_{D}^{2}/{\sigma}_{H}^{2}}} ) + {\sigma}_{H}^{2} |}{ | {\sigma}_{H}^{2}|} \label{formula:mutual_information_4}
\end{align}
\end{subequations}

%-------------------------------------------------------------------------
\section{Reference Guided CDI (RGCDI)}
\label{sec:rgcdi}

\begin{figure}[h]
\begin{subfigure}{.85\linewidth}
    \centering
    \includegraphics[width=85mm]{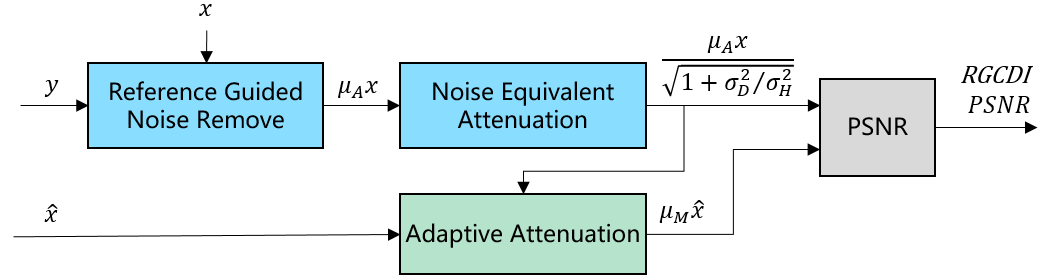}
    \caption{Reference Guided CDI}
    \label{fig:CDI_a}
\end{subfigure}
\qquad

\begin{subfigure}{.85\linewidth}
    \centering
    \includegraphics[width=85mm]{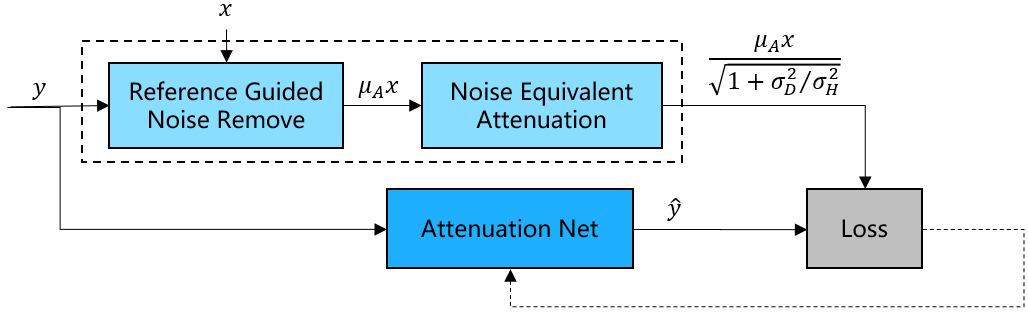}
    \caption{Attenuation Net Training}
    \label{fig:CDI_b}
\end{subfigure}
\qquad

\begin{subfigure}{.85\linewidth}
    \centering
    \includegraphics[width=85mm]{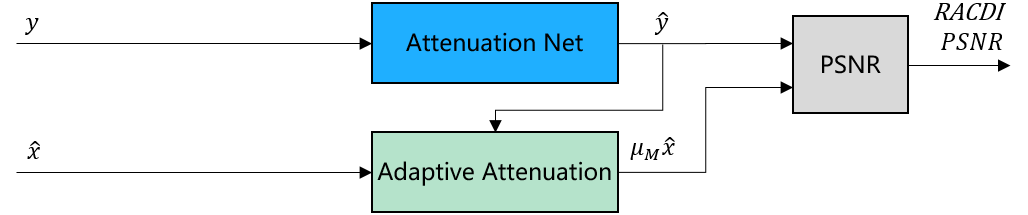}
    \caption{Reference Agnostic CDI}
    \label{fig:CDI_c}
\end{subfigure}

  \caption{(a) Reference Guided CDI calculates the wavelet domain attenuations of reference image and restored image, and obtains $RGCDI\_PSNR$ by comparing the attenuated images. (b) Attenuation Net is trained to predict the attenuated reference image from the degraded image. (c) Reference Agnostic CDI calculates $RACDI\_PSNR$ without reference images.}
  \label{fig:CDI}
\end{figure}

Based on the above fidelity subjective evaluation model, we design a CDI evaluation scheme that can accommodate noise degradations. In this section, we propose the wavelet domain Reference-Guided CDI algorithm, as Fig.\ref{fig:CDI_a} illustrates. The core idea is to remove random noise from the degraded image using the reference image guidance and obtain a noise-free attenuated reference image. Next, noise-equivalent attenuation is applied to further attenuate the image, simulating the impact of noise on the perception of fidelity. Finally, the restored image is adaptively attenuated to match the attenuated reference image as closely as possible and is then compared with it to compute the CDI score.

The wavelet coefficients of the reference image, degraded image and restored image are denoted as $x$, $y$ and $\hat{x}$, respectively. The Reference Guided Noise Remove module uses noise splitting in Sec.\ref{subsec:two_step_degrade} to remove noise from $y$, and obtain an attenuated reference image ${\mu}_{A}x$. The Noise Equivalent Attenuation module (Sec.\ref{subsec:hvs_model}) further attenuates ${\mu}_{A}x$ to ${\mu}_{A}x / \sqrt{1 + {\sigma}_{D}^{2} / {\sigma}_{H}^{2}}$. The Adaptive Attenuation module adaptively attenuates $\hat{x}$ to best match ${\mu}_{A}x / \sqrt{1 + {\sigma}_{D}^{2} / {\sigma}_{H}^{2}}$, and obtains ${\mu}_{M}\hat{x}$. Finally, we can calculate the PSNR and get the $RGCDI\_PSNR$ score.

\subsection{Adaptive Attenuation}

In order to compare the restored image $\hat{x}$ with the attenuated reference image ${\mu}_{A}x / \sqrt{1 + {\sigma}_{D}^{2} / {\sigma}_{H}^{2}}$, $\hat{x}$ should be attenuated accordingly. Considering that BIR degradation is indeterminate (Fig.\ref{fig:degradation_Indeterminacy}), setting the attenuation coefficient of $\hat{x}$ to ${\mu}_{A} / \sqrt{1 + {\sigma}_{D}^{2} / {\sigma}_{H}^{2}}$ is unreasonable. A feasible method is to use the minimization of the squared error (as Eq.\ref{formula:WAM_1} shows) to adaptively calculate the optimal attenuation coefficient $\mu_{M}$.

\begin{equation}\label{formula:WAM_1}
\begin{aligned}
    \mu_{M} &= \arg \min_{\mu_{M}} \left \| \mu_{M}\hat{x} - {\mu}_{A}x / \sqrt{1 + {\sigma}_{D}^{2} / {\sigma}_{H}^{2}} \right\| \\
    &= COV ( \hat{x},\frac{{\mu}_{A}x}{\sqrt{1 + {\sigma}_{D}^{2} / {\sigma}_{H}^{2}}} ) / COV\left(\hat{x}, \hat{x}\right)
\end{aligned}
\end{equation}

Subsequently, the $RGCDI\_PSNR$ score can be computed as $PSNR(IDWT(\mu_{M}x), IDWT(\frac{{\mu}_{A}x}{\sqrt{1 + {\sigma}_{D}^{2}/{\sigma}_{H}^{2}}}) )$, where IDWT is Inverse Discrete Wavelet Transform.

\subsection{RGCDI Property Analysis}

In this section, we combine the above Reference Guided Noise Remove and Noise Equivalent Attenuation as a function $F$), where $F( x, y ) = {\mu}_{A}x / \sqrt{1 + {\sigma}_{D}^{2} / {\sigma}_{H}^{2}}$. The properties of RGCDI are illustrated as follows.

\noindent \textbf{Idempotency.} Multiple $F$ operations are equivalent to a single operation.

\begin{equation}\label{formula:property_1} 
\begin{aligned}
    F(x, y) = F( x, F(x, y) ) = F( F(x, y), y )
\end{aligned}
\end{equation}

\noindent \textbf{Cascade Degradation Property.} The $F$ function of cascading multiple degradations is equivalent to cascading multiple $F$ functions of single degradation.

\begin{equation}\label{formula:property_2}
\begin{aligned}
    F ( x, D_{2} ( D_{1} (x))) = F(F(x, D_{1}(x)), D_{2} ( D_{1} (x)))
\end{aligned}
\end{equation}

\noindent \textbf{$\mathbf{RGCDI\_PSNR \ge PSNR}$ (Conditional).} In general, the attenuation coefficients satisfy $0 \le {\mu}_{A} / \sqrt{1 + {\sigma}_{D}^{2} / {\sigma}_{H}^{2}} \le 1 $. Under this condition, $RGCDI\_PSNR$ is greater than $PSNR$.

\subsection{Estimation of HVS noise parameter ${\sigma}_{H}$}

The noise intensity parameter ${\sigma}_{H}$ is the only hyper parameter that requires manual adjustment. Similarly to \cite{VIF}, ${\sigma}_{H}$ depends on the dynamic range of the input signal and can be adjusted by $\lambda$, as illustrated in Eq.\ref{formula:hvs_noise}. The experiments in Fig.\ref{fig:lambda} show that the accuracy of RGCDI is not sensitive to changes in $\lambda$.

\begin{equation}\label{formula:hvs_noise}
\begin{aligned}
    \sigma_H^2 = \lambda \cdot COV( \mu_A x, \mu_A x ) = \lambda \cdot \mu_A^2 \cdot COV( x, x )
\end{aligned}
\end{equation}

\section{Reference Agnostic CDI (RACDI)}
\label{sec:racdi}

The RGCDI proposed in the previous section depends on the reference images. In this section, we extend the RGCDI to a Reference-Agnostic CDI algorithm that does not require reference images and can achieve the CDI calculation for real-world BIR. By observation, it is evident that the Reference Guided Noise Removal operation eliminates noise from the degraded image while retaining the attenuation. The Noise Equivalent Attention operation further attenuates the image according to the noise. The output image ${\mu}_{A}x / \sqrt{1 + {\sigma}_{D}^{2} / {\sigma}_{H}^{2}}$ is actually a denoised and blurred version of $y$. This observation inspires us to adopt a denoising model to directly convert $y$ into ${\mu}_{A}x / \sqrt{1 + {\sigma}_{D}^{2} / {\sigma}_{H}^{2}}$.

As Fig.\ref{fig:CDI_b} shows, paired ($y$, ${\mu}_{A}x / \sqrt{1 + {\sigma}_{D}^{2} / {\sigma}_{H}^{2}}$) training images of various types of degradation can be generated through the RGCDI system. After end-to-end training, the Attenuation Net is able to predict ${\mu}_{A}x / \sqrt{1 + {\sigma}_{D}^{2} / {\sigma}_{H}^{2}}$ from $y$. Many off-the-shelf denoise models, such as SCUNet \cite{SCUNet} and Restormer \cite{Restormer} \textit{et al.}, can be used as the candidate Attenuation Net backbone. In Fig.\ref{fig:CDI_c}, the output of Attenuation Net is denoted as $\hat{y}$. We replace ${\mu}_{A}x / \sqrt{1 + {\sigma}_{D}^{2} / {\sigma}_{H}^{2}}$ of the RGCDI structure with $\hat{y}$, so that the $RACDI\_PSNR$ scores can be calculated.

\begin{figure}[thb]
  \centering
  \includegraphics[width=85mm]{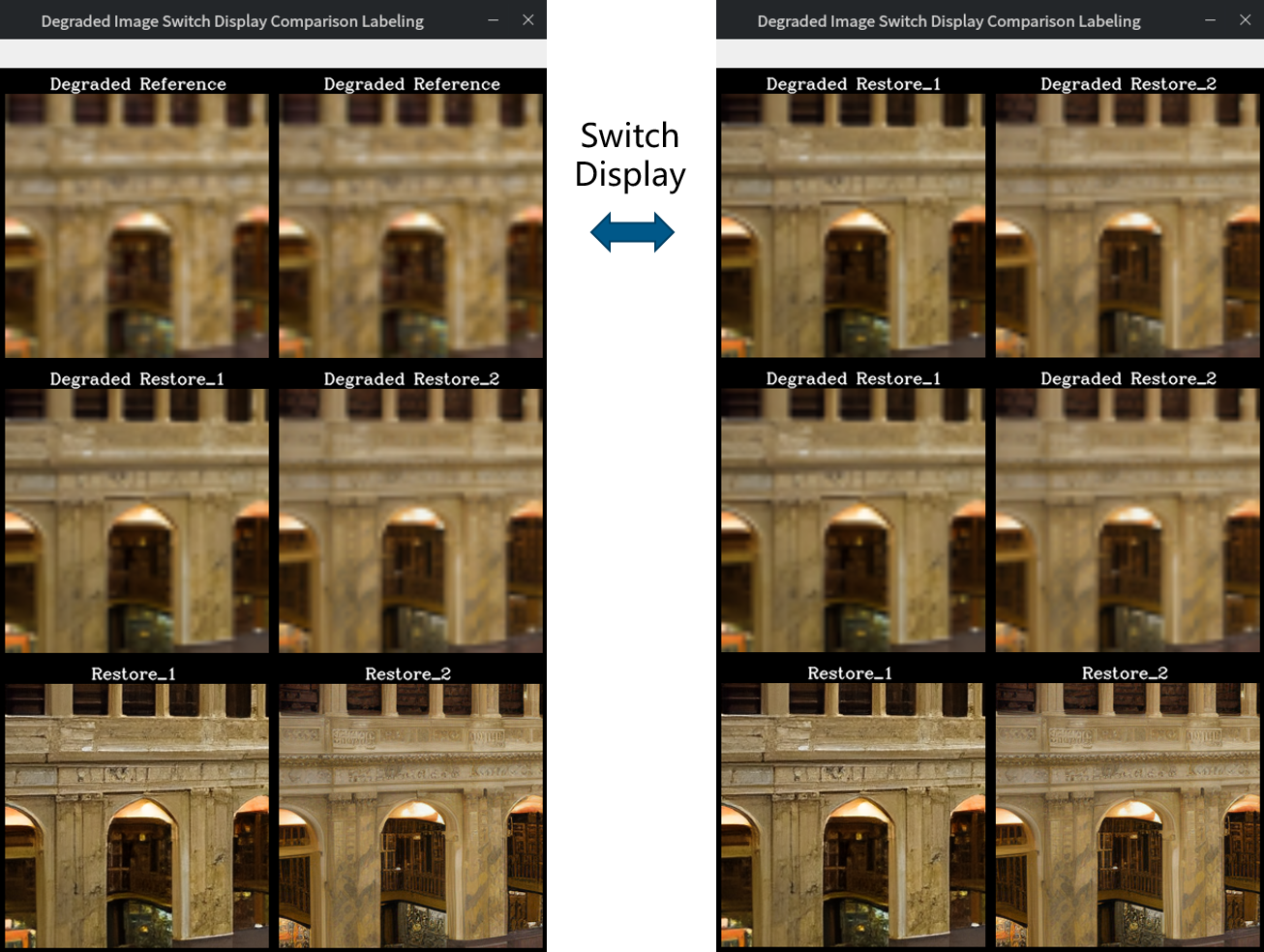}
  \caption{DISDCD annotation interface}
  \label{fig:DISDCD}
\end{figure}

\section{Degraded Images Switch Display Comparison Dataset (DISDCD)}
\label{sec:dataset}

The existing IQA datasets \cite{LIVE,CSIQ,TID2008,TID2013,PieAPP,BAPPS,QADS,PIPAL} fail to assess the performance of different IQA methods in terms of fidelity evaluation. In this section, we build the first IQA dataset with fidelity evaluation based on degraded images, named the Degraded Images Switch Display Comparison Dataset (DISDCD). It contains restored images corresponding to 4 different types of degradation, including down sampling (4x), Gaussian noise ($\sigma\!=\!50$), JPEG (QF10) and combined degradation (Real-ESRGAN second-order degradation \cite{RealESRGAN}). Each degradation type contains 100 degraded images from DIV2K \cite{DIV2K}, and 200 restored images. BIR restoration algorithms include BSRGAN \cite{BSRGAN}, Real-ESRGAN \cite{RealESRGAN}, LDL \cite{LDL}, DASR \cite{DASR}, FeMaSR \cite{FeMaSR}, LDM \cite{LDM}, StableSR \cite{StableSR}, ResShift \cite{ResShiftED}, PASD \cite{PASD}, DiffBIR \cite{DiffBIR} and SeeSR \cite{SeeSR}. 

The conventional method of comparing left and right images makes it difficult to distinguish fidelity differences. To address this issue, the proposed DISDCD utilizes the double-stimulus continuous quality scale method of ITU-R BT.500-15 \cite{BT500} for subjective judgment, which helps raters more accurately identify fidelity differences. The Two-Alternative Forced Choice (2AFC) \cite{LPIPS} method is adopted to derive the subjective fidelity score. During dataset creation, we first select a reference image and apply a degradation operation to obtain the degraded image. We then randomly select two different BIR algorithms to generate two restored images. To better observe the differences between the degraded images, we apply the same degradation operation on the two restored images, producing the degraded restored images. Then a rater is asked to determine which is closer to the degraded image, and the response is recorded. Note that when generating Gaussian noise during the degradation process, a fixed random seed is required to maintain unchanged noise.

Fig.\ref{fig:DISDCD} illustrates the DISDCD annotation interface. The top row presents two identical degraded images. The middle row displays the two degraded restored images, and the bottom row displays the two original restored images. Raters can use buttons to toggle between the degraded image and degraded restored images presented in the top row as many times as needed, until they make a clear judgment.

\section{Experiments}
\label{sec:experiments}

In this section, we conduct experiments to verify the effectiveness of the proposed RGCDI and RACDI methods. Since the existing FR-IQA-based dataset cannot be used for BIR fidelity evaluation, we do comparative experiments on our proposed DISDCD dataset. Sec.\ref{subsec:comparison_disdcd} compares the proposed methods with the common FR-IQA methods. Sec.\ref{subsec:backbones_selection} compares 4 candidate Attenuation Net backbones in 7 types of image restoration tasks. Sec.\ref{subsec:racdi_test} tests the errors of $RACDI\_PSNR$ relative to $RGCDI\_PSNR$.

\begin{figure}[htb]
  \centering
  \includegraphics[width=85mm]{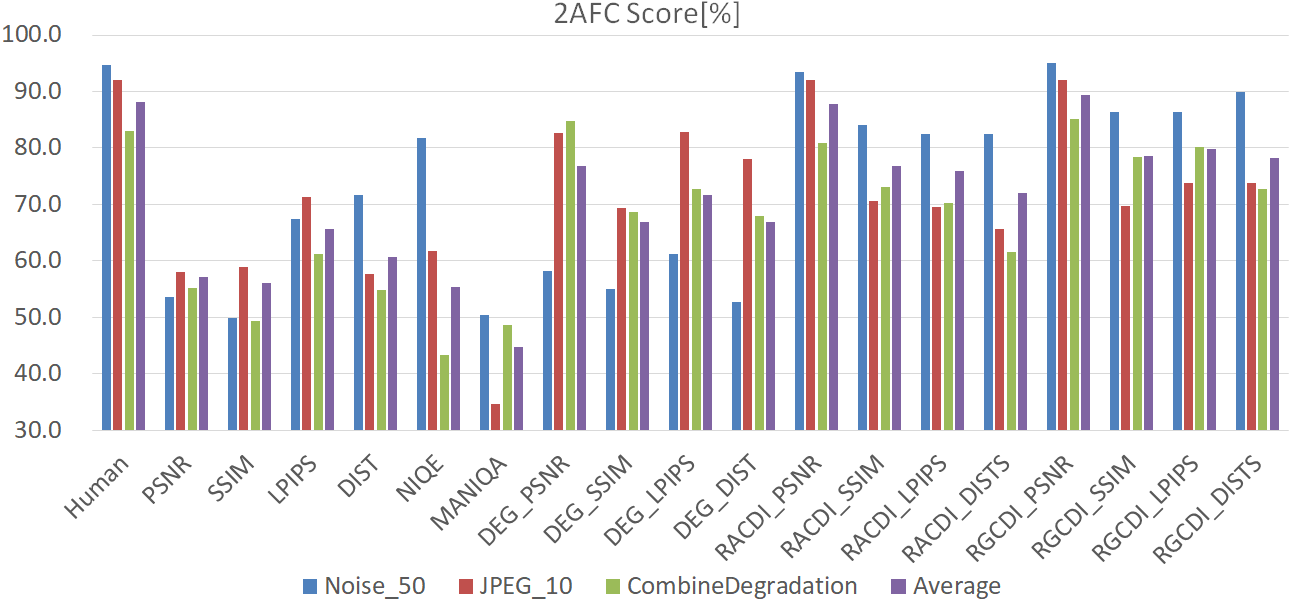}
  \caption{Evaluation on DISDCD. The proposed $RGCDI\_PSNR$ and $RACDI\_PSNR$ have significantly higher 2AFC scores, close to human judgments.}
  \label{fig:2AFC}
\end{figure}

\begin{figure}[htb]
  \centering
  \includegraphics[width=40mm]{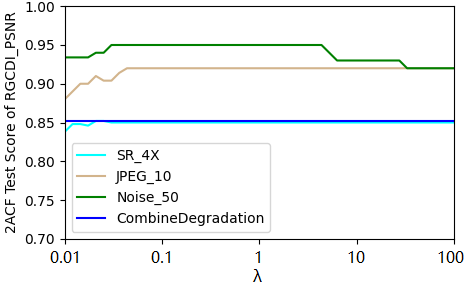}
  \caption{Hyper parameter ($\lambda$) tuning.}
  \label{fig:lambda}
\end{figure}

\begin{figure}[htb]
  \centering
  \includegraphics[width=75mm]{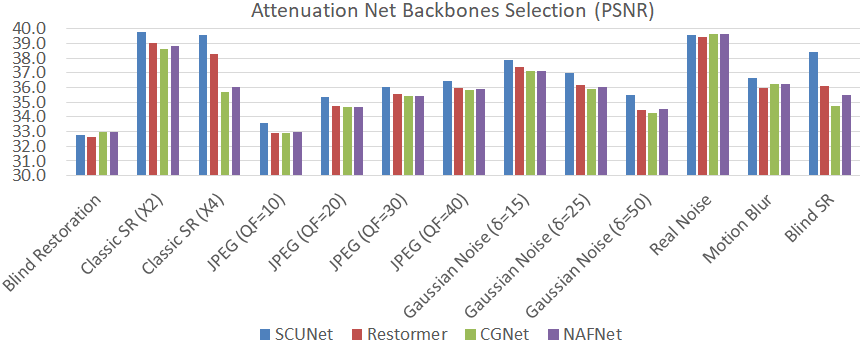}
  \caption{Attenuation Net Backbones Comparison.}
  \label{fig:backbone_selection}
\end{figure}

\begin{table*}[thb]
    \centering
    \resizebox{1.0\textwidth}{!}{
    \begin{tabular}{>{\centering\arraybackslash}p{1.33cm}|>{\centering\arraybackslash}p{1.1cm}|>{\centering\arraybackslash}p{2.7cm}|>{\centering\arraybackslash}p{2.7cm}|>{\centering\arraybackslash}p{2.7cm}|>{\centering\arraybackslash}p{2.75cm}|>{\centering\arraybackslash}p{2.7cm}|>
    {\centering\arraybackslash}p{2.7cm}}
    \hline
         \multicolumn{2}{c|}{} & MT-RNN \cite{MT-RNN} & DMPHN \cite{DMPHN} & MIMO-UNet+ \cite{MIMO-UNet+} & MPRNet \cite{MPRNet} & Restormer \cite{Restormer} & GRL-B \cite{GRL} \\
    \hline
         \multicolumn{2}{c|}{\textbf{Motion Deblur}} & 38.03 / 1.62 & 38.10 / 1.58 & 38.04 / 1.68 & 37.89 / 1.78 & 37.95 / 1.74 & 37.83 / 1.94 \\
    \hline
    \hline
         \multicolumn{2}{c|}{} & RCAN \cite{RCAN} & SAN \cite{SAN} & HAN \cite{HAN} & IPT \cite{IPT} & SwinIR \cite{SwinIR} & GRL-S \\
    \hline
         \multirow{2}{*}{\textbf{ClassicSR}} & x2 & 39.78 / 0.23 & 39.80 / 0.20 & 39.79 / 0.22 & 39.69 / 0.31 & 39.75 / 0.22 & 39.70 / 0.30 \\
    \cline{2-8}
         & x4 & 39.57 / 0.32 & 39.62 / 0.28 & 39.58 / 0.40 & 39.41 / 0.55 & 39.50 / 0.41 & 39.37 / 0.60 \\
    \hline
    \hline
         \multicolumn{2}{c|}{} & DnCNN \cite{DNCNN} & IPT & DRUNet \cite{DRUNet} & SwinIR & Restormer & SCUNet \cite{SCUNet} \\
    \hline
         \multirow{3}{*}{\makecell{\textbf{Gaussian}\\
         \textbf{Denoise}}} & $\sigma=15$ & 38.20 / 1.97 & - & 39.02 / 1.28 & 39.63 / 1.66 & 38.97 / 1.10 & 39.03 / 1.11 \\         
    \cline{2-8}
         & $\sigma=25$ & 37.42 / 1.99 & - & 38.63 / 1.49 & 39.44 / 2.17 & 38.49 / 1.28 & 38.53 / 1.28 \\
    \cline{2-8}
         & $\sigma=50$ & 35.63 / 1.57 & 34.31 / 1.91 & 37.55 / 1.58 & 39.03 / 2.74 & 37.52 / 1.39 & 37.55 / 1.35 \\
    \hline
    \hline
         \multicolumn{2}{c|}{} & AT-BSN(D) \cite{AT-BSN} & S-Adaptive \cite{S-Adaptive} & SDAP \cite{SDAP} & InvDN \cite{InvDN} & DANet \cite{DANet} & VDN \cite{VDN}\\
    \hline
         \multicolumn{2}{c|}{\textbf{Real Denoise}} & 39.27 / 0.62 & 39.35 / 0.40 & 39.14 / 0.70 & 39.91 / 0.14 & 39.91 / 0.13 & 39.91 / 0.13 \\
    \hline
    \hline
         \multicolumn{2}{c|}{} & SeeSR \cite{SeeSR} & DiffBIR \cite{DiffBIR} & StableSR \cite{StableSR} & RealESRGAN \cite{RealESRGAN} & BSRGAN \cite{BSRGAN} & FeMaSR \cite{FeMaSR}\\
    \hline
         \multicolumn{2}{c|}{\textbf{Blind Restoration}} & 33.29 / 1.79 & 35.51 / 2.48 & 30.95 / 1.31 & 32.38 / 1.70 & 34.83 / 2.64 & 34.82 / 2.57 \\
    \hline
    \hline
         \multicolumn{2}{c|}{} & QGAC \cite{QGAC} & FBCNN \cite{FBCNN} & GRL-S & - & - & - \\
    \hline
         \multirow{4}{*}{\makecell{\textbf{JPEG}\\
         \textbf{Artifact}\\ \textbf{Reduction}}} & QF10 & 39.52 / 5.56 & 39.94 / 5.79 & 39.93 / 5.15 & - & - & - \\
    \cline{2-8}
         & QF20 & 39.83 / 4.12 & 39.99 / 4.09 & 39.98 / 3.57 & - & - & - \\
    \cline{2-8}
         & QF30 & 39.86 / 3.36 & 39.98 / 3.34 & 39.97 / 2.81 & - & - & - \\	
    \cline{2-8}
         & QF40 & 39.89 / 2.95 & 39.99 / 2.94 & 39.98 / 2.40 & - & - & - \\		
    \hline
    \end{tabular}
    }
    \caption{\textbf{Comparison between $RACDI\_PSNR$ and $RGCDI\_PSNR$}: For each cell of data in this table, the number on the left represents RACDI\_PSNR and the number on the right represents the mean error of RACDI\_PSNR relative to RGCDI\_PSNR. The following tasks and their respective results are presented: (1) Motion deblur task uses GoPro dataset, (2) ClassSR task uses Urban100 dataset, (3) Gaussian denoise task uses Urban100 dataset (Note: Official IPT model does not support $\sigma$=15, 25), (4) Real deoise task uses SIDD dataset, (5) Blind restoration task uses DIV2K dataset, (6) JPEG compression artifact reduction task uses BSDS500 dataset.}
    \label{table_algorithm_psnr}
\end{table*}

\subsection{Comparison of IQA Methods on DISDCD}
\label{subsec:comparison_disdcd}

This experiment compares the proposed RGCDI and RACDI with common FR-IQA methods (including PSNR, SSIM \cite{SSIM}, LPIPS \cite{LPIPS}, DISTS \cite{DISTS}), FR-IQA on degraded images and human judgments. $DEG\_PSNR$ represents PSNR on degraded images (Sec.\ref{sec:friqa_on_degraded}). The DISDCD validation sets contain 5 pairwise judgments for each sample. Following the approach in \cite{LPIPS}, human accuracy is computed as shown in Eq.\ref{formula:Human_score}, where $p$ represents subjective judgment similarity. For example, if there are 4 preferences for one restored image and 1 for the other, $p=0.8$. 

\begin{equation}\label{formula:Human_score}
    2AFC\_Human = p^{2} + \left (1 - p \right )^{2}
\end{equation}

Similar to \cite{LPIPS}, we calculated the average 2AFC scores of all subjective judgments. As Fig.\ref{fig:2AFC} illustrates, the 2AFC scores of PSNR, SSIM, LPIPS and DISTS are all obviously lower. The $DEG\_PSNR$ achieves higher scores, except for Gaussian noise and JPEG, where it faces challenges in noise processing. The proposed $RGCDI\_PSNR$ achieves the highest score, with some scores even exceeding human scores.
$RACDI\_PSNR$ is slightly lower than $RGCDI\_PSNR$. The low scores of the NR-IQA algorithms (NIQE, MANIQA) are due to the fact that they cannot evaluate fidelity.

\noindent \textbf{Hyper Parameter ($\lambda$) Tunning. }  As Fig.\ref{fig:lambda} illustrates, the 2AFC scores of $RGCDI\_PSNR$ on DISDCD are not sensitive to changes of $\lambda$, with a recommended value of 0.3.

\noindent \textbf{CDI Calculation Efficiency.} The main computational load of RGCDI lies in the Discrete Fourier Transform, while for RACDI, it is in SCUNet. The test times for $RGCDI\_PSNR$ and $RACDI\_PSNR$ are 48 ms and 21 ms, respectively. The testing conditions are as follows: image resolution 512x512, using an A10 GPU, inference performed directly through PyTorch. The computational efficiency meets the requirements for large-batch evaluation.

\subsection{Attenuation Net Backbones Selection}
\label{subsec:backbones_selection}
We train a general Attenuation Net on various IR tasks simultaneously. Specifically, we select 7 types of degradation including combined degradation \cite{RealESRGAN}, down sampling, JPEG compression, Gaussian noise, real noise, motion blur, and blind SR degradation. We select 4 types of off-the-shelf De-noise networks (SCUNet \cite{SCUNet}, Restormer \cite{Restormer}, CGNet \cite{CGNet} and NAFNet \cite{NAFNet}) as candidate backbones. Each model was trained for 200k iterations and then tested on various validation sets. As the experimental results in Fig.\ref{fig:backbone_selection} show, SCUNet achieves the best performance, followed by Restormer in second place, while the performance of CGNet and NAFNet is relatively lower.

\subsection{RACDI Testing}
\label{subsec:racdi_test}
In this experiment, we use SCUNet as the backbone and test $RACDI\_PSNR$ on 6 types of image restoration tasks (classic SR, motion deblur, Gaussian denoise, real denoise, JPEG compression artifact reduction, and blind restoration \cite{RealESRGAN}). Table.\ref{table_algorithm_psnr} shows that the mean errors of motion deblur, classic SR, Gaussian denoise, real denoise and blind restoration are less than 2.74dB. The mean errors of JPEG restoration are higher, with a maximum of 5.79dB.

\section{Conclusion}
\label{sec:conclusion}

In this paper, we propose the development of a new CDI-based IQA framework, and introduce a wavelet-based reference guided CDI algorithm, which improves the accuracy of evaluating the consistency of degraded images. Furthermore, we train a general attenuation net to realize a reference-agnostic CDI calculation. Finally, we create a new Degraded Images Switch Display Comparison Dataset. Experiments demonstrate that the proposed CDI algorithm achieves the highest score and some scores even exceed human judgment. We anticipate that CDI will become a benchmark for fidelity evaluation and will contribute to guiding blind image restoration algorithms to improve fidelity.

{
    \small
    \bibliographystyle{ieeenat_fullname}
    \bibliography{main}
}

\clearpage
\setcounter{page}{1}
\maketitlesupplementary

\section{Image Restoration Algorithm of Fig.\ref{fig:resolution_non-unique}}

\noindent{Given}:

\begin{equation}\label{formula:supplementary_restoration_1_1} 
\begin{aligned}
    &k = Kernel\left( size=9\times9,\sigma=5 \right) \\
    &I_{y} = conv \left( I_{x}, k \right) \\
\end{aligned}
\end{equation}

\noindent{Using existing BIR algorithms to restore $I_{x}$}:

\begin{equation}\label{formula:supplementary_restoration_1_2} 
    I_{\hat{x}0} = BIR\left( I_{y} \right)
\end{equation}

\noindent{Design Loss function}:

\begin{equation}\label{formula:supplementary_restoration_1_3} 
    L = \|  conv ( I_{\hat{x}}, k ) - I_{y} \| + \lambda \|  I_{\hat{x}} - I_{\hat{x}0} \|
\end{equation}

\noindent{Set $I_{\hat{x}0}$ as the initial value of $I_{\hat{x}}$. Use gradient descent to minimize $L$ and update $I_{\hat{x}}$. In Eq.\ref{formula:supplementary_restoration_1_3}, $\lambda \|  I_{\hat{x}} - I_{\hat{x}0} \|$ is the regularization term, $\lambda \in [0.001,0.01]$.}

\section{Image Restoration Algorithm of Fig.\ref{fig:degradation_Indeterminacy}}

\noindent{Given}:

\begin{equation}\label{formula:supplementary_restoration_2_1} 
\begin{aligned}
    &k_{1} = Kernel\left( size=13\times13,\sigma=1.0 \right) \\
    &k_{2} = Kernel\left( size=13\times13,\sigma=1.06, \enspace 1.12 \enspace  or \enspace  1.18 \right) \\
    &I_{y} = conv \left( I_{x}, k_{1} \right)
\end{aligned}
\end{equation}

\noindent{Design Loss function}:

\begin{equation}\label{formula:supplementary_restoration_2_2} 
    L = \|  conv ( I_{\hat{x}}, k_{2} ) - I_{y} \| + \lambda \|  I_{\hat{x}} - I_{x} \|
\end{equation}

\noindent{Set $I_{x}$ as the initial value of $I_{\hat{x}}$. Use gradient descent to minimize $L$ and update $I_{\hat{x}}$. In Eq.\ref{formula:supplementary_restoration_2_2}, $\lambda \|  I_{\hat{x}} - I_{x} \|$ is the regularization term, $\lambda \in [0.001,0.01]$.}

\section{Proof of RGCDI Property}

\noindent \textbf{Idempotency.} Multiple $F$ operations are equivalent to a single operation, $F( x, y ) = \frac{{\mu}_{A}x}{\sqrt{1 + {\sigma}_{D}^{2} / {\sigma}_{H}^{2}}}$.

\noindent{Given}:

\begin{equation}\label{formula:supplementary_property_1_1} 
\begin{aligned}
    &{\mu}_{A} = COV \left( y, x \right) / COV \left ( x,x \right ) \\
    &{n}_{D} = y - {\mu}_{A}x\\
    &\sigma_{D}^{2} = COV \left (y , y \right ) - {\mu}_{A} COV \left (y , x \right )
\end{aligned}
\end{equation}

\noindent{Calculating $F( x, F( x, y ) )$}:

\begin{equation}\label{formula:supplementary_property_1_2} 
\begin{aligned}
    &{\mu'}_{A} = COV(F(x,y), x)/COV(x, x) \\
    &{\mu'}_{A} = COV(\frac{{\mu}_{A}x}{\sqrt{1 + {\sigma}_{D}^{2} / {\sigma}_{H}^{2}}}, x)/COV(x, x) \\
    &{\mu'}_{A} = \frac{{\mu}_{A}}{\sqrt{1 + {\sigma}_{D}^{2} / {\sigma}_{H}^{2}}} \\
    &{n'}_{D} = F(x,y) - {\mu'}_{A}x = 0 \\
    &{\sigma'}_{D}^{2} = 0 \\
    &F( x, F( x, y ) ) = \frac{{\mu'}_{A}x}{\sqrt{1 + {\sigma'}_{D}^{2} / {\sigma}_{H}^{2}}} \\
    &F( x, F( x, y ) ) = {\mu}'_{A}x =  \frac{{\mu}_{A}x}{\sqrt{1 + {\sigma}_{D}^{2} / {\sigma}_{H}^{2}}} \\
    &F( x, F( x, y ) ) = F( x, y )
\end{aligned}
\end{equation}

\begin{table*}[thb]
    \centering
    \begin{tabular}{c|>{\centering\arraybackslash}p{1.1cm}|>{\centering\arraybackslash}p{1.1cm}|>{\centering\arraybackslash}p{1.1cm}|>{\centering\arraybackslash}p{1.1cm}|>{\centering\arraybackslash}p{1.1cm}|>{\centering\arraybackslash}p{1.1cm}|>{\centering\arraybackslash}p{1.1cm}|>{\centering\arraybackslash}p{1.1cm}|>{\centering\arraybackslash}p{1.1cm}|>{\centering\arraybackslash}p{1.1cm}}
    \hline
         \multirow{2}{*}{\parbox[c]{\widthof{Method}}{\centering Method}} & \multicolumn{2}{c|}{\textbf{Set5}} & \multicolumn{2}{c|}{\textbf{Set14}} & \multicolumn{2}{c|}{\textbf{BSD100}} & \multicolumn{2}{c|}{\textbf{Manga109}} & \multicolumn{2}{c}{\textbf{Urban100}}\\
         \cline{2-11}
         & x2 & x4 & x2 & x4 & x2 & x4 & x2 & x4 & x2 & x4\\
    \hline
         SCUNet & \best{40.00} & \best{39.87} & \best{39.70} & \best{39.44} & \best{39.84} & \best{39.75} & \best{39.85} & \best{39.47} & \best{39.30} & \best{39.10} \\ 
    \hline
         Restormer & 38.41 & 36.08 & 39.04 & 37.65 & 39.83 & 39.71 & 38.77 & 39.35 & 39.02 & 38.62 \\
    \hline
         CGNet & 37.54 & 35.42 & 38.35 & 34.36 & 39.77 & 39.69 & 38.74 & 33.02 & 38.77 & 35.87 \\
    \hline
         NAFNet & 38.24 & 36.21 & 38.39 & 34.96 & 39.78 & 39.70 & 39.01 & 33.14 & 38.73 & 36.29 \\
    \hline
    \end{tabular}
    \caption{Attenuation Net backbone comparison on \textbf{ClassicSR} validation sets $\left( PSNR \right)$: SCUNet is obviously better than others.}
    \label{table_ClassicSR_psnr}
\end{table*}

\begin{table*}[thb]
    \centering
    \resizebox{1.0\textwidth}{!}{
    \begin{tabular}{c|>{\centering\arraybackslash}p{1.04cm} >{\centering\arraybackslash}p{1.04cm} >{\centering\arraybackslash}p{1.04cm}|>{\centering\arraybackslash}p{1.04cm} >{\centering\arraybackslash}p{1.04cm} >{\centering\arraybackslash}p{1.04cm}|>{\centering\arraybackslash}p{1.04cm} >{\centering\arraybackslash}p{1.04cm} >{\centering\arraybackslash}p{1.04cm}|>{\centering\arraybackslash}p{1.04cm} >{\centering\arraybackslash}p{1.04cm} >{\centering\arraybackslash}p{1.04cm}|>{\centering\arraybackslash}p{1.04cm}}
    \hline
         \multirow{2}{*}{\parbox[c]{\widthof{Method}}{\centering Method}} & \multicolumn{3}{c|}{\textbf{CBSD68}} & \multicolumn{3}{c|}{\textbf{Kodak24}} & \multicolumn{3}{c|}{\textbf{McMaster}} & \multicolumn{3}{c|}{\textbf{Urban100}} & \multirow{2}{*}{\parbox[c]{\widthof{Method}}{\centering \textbf{SIDD}}}\\
         \cline{2-13}
         & $\sigma=15$ & $\sigma=25$ & $\sigma=50$ & $\sigma=15$ & $\sigma=25$ & $\sigma=50$ & $\sigma=15$ & $\sigma=25$ & $\sigma=50$ & $\sigma=15$ & $\sigma=25$ & $\sigma=50$ &\\
    \hline
         SCUNet & \best{37.60} & \best{36.63} & \best{34.80} & \best{38.63} & 37.90 & 36.70 & \best{38.49} & \best{37.56} & \best{35.82} & \best{36.62} & \best{35.80} & \best{34.63} & \best{39.57} \\
    \hline
         Restormer & 36.86 & 35.26 & 33.19 & \best{38.63} & \best{37.92} & 36.70 & 37.37 & 35.71 & 33.59 & 36.60 & 35.71 & 34.41 & 39.43 \\
    \hline
         CGNet & 36.70 & 34.79 & 32.20 & 38.51 & 37.84 & 36.84 &	37.18 & 35.75 & 33.82 &	36.09 & 35.31 & 34.14 & 34.25 \\
    \hline
         NAFNet & 36.60 & 34.99 & 33.03 & 38.52 & 37.86 & \best{36.86} & 37.28 & 35.78 & 33.86 & 36.14 & 35.36 & 34.23 & 34.50 \\
    \hline
    \end{tabular}
    }
    \caption{Attenuation Net backbone comparison on \textbf{Gaussian De-noise} validation sets $\left( PSNR \right)$: SCUNet and Restormer are slightly better than CGNet and NAFNet.}
    \label{table_Noise_psnr}
\end{table*}

\begin{table}[thb]
    \centering
    \resizebox{0.50\textwidth}{!}{
    \begin{tabular}{c|>{\centering\arraybackslash}p{1.3cm}|>{\centering\arraybackslash}p{1.3cm}|>{\centering\arraybackslash}p{1.3cm}|>{\centering\arraybackslash}p{1.3cm}}
    \hline
         Method & SCUNet & Restormer & CGNet & NAFNet\\
    \hline
         \textbf{GoPro} & 36.69 & 36.13 & 37.07 & \best{37.09} \\
    \hline
         \textbf{HIDE} & \best{37.74} & 36.62 & 37.51 & 37.32 \\
    \hline
         \textbf{RealBlur-J} & \best{34.20} & 33.12 & 32.55 & 32.73 \\
    \hline
         \textbf{RealBlur-R} & \best{38.01} & 37.90 & 37.78 & 37.86 \\
    \hline
    \end{tabular}
    }
    \caption{Attenuation Net backbone comparison on \textbf{MotionBlur} validation sets $\left( PSNR \right)$: SCUNet and Restormer are slightly better than others.}
    \label{table_MotionBlur_psnr}
\end{table}

\begin{table}[thb]
    \centering
    \resizebox{0.50\textwidth}{!}{
    \begin{tabular}{p{1.1cm}|c|>{\centering\arraybackslash}p{1.3cm}|>{\centering\arraybackslash}p{1.3cm}|>{\centering\arraybackslash}p{1.3cm}|>{\centering\arraybackslash}p{1.3cm}}
    \hline
         \multicolumn{2}{c|}{Method} & SCUNet & Restormer & CGNet & NAFNet\\
    \hline
         \multirow{4}{*}{\textbf{LIVE1}} & QF10 & \best{33.19} & 33.13 & 33.09 & 33.16 \\
    \cline{2-6}
         & QF20 & \best{34.99} & 34.85 & 34.74 & 34.79 \\
    \cline{2-6}
         & QF30 & \best{35.71} & 35.57 & 35.41 & 35.47 \\
    \cline{2-6}
         & QF40 & \best{36.11} & 35.97 & 35.77 & 35.83 \\
    \hline
         \multirow{4}{*}{\textbf{BSD500}} & QF10 & \best{34.00} & 32.73 & 32.71 & 32.73 \\
    \cline{2-6}
         & QF20 & \best{35.71} & 34.63 & 34.56 & 34.60 \\
    \cline{2-6}
         & QF30 & \best{36.39} & 35.51 & 35.39 & 35.43 \\
    \cline{2-6}
         & QF40 & \best{36.77} & 36.01 & 35.87 & 35.89 \\
    \hline
    \end{tabular}
    }
    \caption{Attenuation Net backbone comparison on \textbf{JPEG Restoration} validation sets $\left( PSNR \right)$: SCUNet is slightly better than other models.}
    \label{table_JPEG_psnr}
\end{table}

\begin{table}[thb]
    \centering
    \resizebox{0.50\textwidth}{!}{
    \begin{tabular}{>{\centering\arraybackslash}p{1.1cm}|c|>{\centering\arraybackslash}p{1.3cm}|>{\centering\arraybackslash}p{1.3cm}|>{\centering\arraybackslash}p{1.3cm}|>{\centering\arraybackslash}p{1.3cm}}
    \hline
         \multicolumn{2}{c|}{Method} & SCUNet & Restormer & CGNet & NAFNet\\
    \hline
         \multirow{2}{*}{\includegraphics[width=0.7cm]{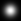}} & $\sigma=5$ & \best{39.29} & 36.43 & 35.49 & 36.31 \\
    \cline{2-6}
         & $\sigma=10$ & \best{38.37} & 35.12 & 34.31 & 35.00 \\
    \hline
         \multirow{2}{*}{\includegraphics[width=0.7cm]{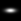}} & $\sigma=5$ & \best{39.12} & 37.27 & 35.24 & 36.42 \\
    \cline{2-6}
         & $\sigma=10$ & \best{37.78} & 35.71 & 34.47 & 35.03 \\
    \hline
         \multirow{2}{*}{\includegraphics[width=0.7cm]{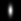}} & $\sigma=5$ & \best{39.23} & 37.28 & 35.21 & 36.22 \\
    \cline{2-6}
         & $\sigma=10$ & \best{37.78} & 35.51 & 34.41 & 34.96 \\
    \hline
         \multirow{2}{*}{\includegraphics[width=0.7cm]{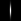}} & $\sigma=5$ & \best{38.52} & 36.45 & 34.56 & 35.49 \\
    \cline{2-6}
         & $\sigma=10$ & \best{37.19} & 34.95 & 34.01 & 34.56 \\
    \hline
    \end{tabular}
    }
    \caption{Attenuation Net backbone comparison on \textbf{BlindSR} validation sets, use Set14 dataset $\left( PSNR \right)$: SCUNet is obviously better than other models.}
    \label{table_Anisotropic_psnr}
\end{table}

\begin{table}[thb]
    \centering
    \resizebox{0.50\textwidth}{!}{
    \begin{tabular}{c|>{\centering\arraybackslash}p{1.3cm}|>{\centering\arraybackslash}p{1.3cm}|>{\centering\arraybackslash}p{1.3cm}|>{\centering\arraybackslash}p{1.3cm}}
    \hline
         Method & SCUNet & Restormer & CGNet & NAFNet \\
    \hline
         \textbf{DIV2K} & 34.90 & 34.97 & 35.58 & \best{35.61} \\
    \hline
         \textbf{CelebA} & 33.95 & 33.84 & \best{34.49} & \best{34.49} \\
    \hline
         \textbf{DrealSR} & \best{31.63} & 31.53 & 31.53 & 31.37 \\
    \hline
         \textbf{RealSR} & \best{30.48} & 30.31 & 30.40 & 30.36 \\
    \hline
    \end{tabular}
    }
    \caption{Attenuation Net backbone comparison on \textbf{Blind Restoration} validation sets $\left( PSNR \right)$: There is no significant difference in the PSNR scores of all models.}
    \label{table_BlindRestoration_psnr}
\end{table}

\begin{table*}[h]
    \centering
    \resizebox{1.0\textwidth}{!}{
    \begin{tabular}{>{\centering\arraybackslash}p{1.5cm}|>{\centering\arraybackslash}p{1.2cm}|>{\centering\arraybackslash}p{2.3cm}|>{\centering\arraybackslash}p{2.3cm}|>{\centering\arraybackslash}p{2.3cm}|>{\centering\arraybackslash}p{2.3cm}|>{\centering\arraybackslash}p{2.3cm}|>
    {\centering\arraybackslash}p{2.3cm}}
    \hline
         \multicolumn{2}{c|}{} & RCAN & SAN & HAN & IPT & SwinIR & GRL-S\\
    \hline
         \multirow{2}{*}{\textbf{BSD100}} & x2 & 40.0 / 0.08	& 40.0 / 0.08 & 40.0 / 0.08 & 40.0 / 0.08 & 40.0 / 0.07 & 40.0 / 0.08 \\
    \cline{2-8}
         & x4 & 39.96 / 0.19 & 39.98 / 0.16 & 39.98 / 0.16 & 39.96 / 0.17 & 39.96 / 0.13 & 39.82 / 0.20 \\
    \hline
         \multirow{2}{*}{\textbf{Manga109}} & x2 & 39.98 / 0.02 & 39.98 / 0.02 & 39.98 / 0.02 & 39.98 / 0.02 & 39.97 / 0.03 & 39.98 / 0.02 \\ 
    \cline{2-8}
         & x4 & 39.83 / 0.12 & 39.84 / 0.10 & 39.84 / 0.10 & 39.81 / 0.12 & 39.77 / 0.16 & 39.80 / 0.15 \\
    \hline
    \end{tabular}
    }
    \caption{RACDI test results, \textbf{Classic SR} on BSD100 and Manga109: For each cell of data in this table, the number on the left represents RACDI\_PSNR and the number on the right represents the mean error of RACDI\_PSNR relative to RGCDI\_PSNR. Maximum Mean Error 0.20dB}
    \label{table_classicsr_cdi}
\end{table*}

\begin{table*}[h]
    \centering
    \resizebox{1.0\textwidth}{!}{
    \begin{tabular}{>{\centering\arraybackslash}p{1.5cm}|>{\centering\arraybackslash}p{1.2cm}|>{\centering\arraybackslash}p{2.3cm}|>{\centering\arraybackslash}p{2.3cm}|>{\centering\arraybackslash}p{2.3cm}|>{\centering\arraybackslash}p{2.3cm}|>{\centering\arraybackslash}p{2.3cm}|>
    {\centering\arraybackslash}p{2.3cm}}
    \hline
         \multicolumn{2}{c|}{} & DnCNN & IPT & DRUNet & SwinIR & Restormer & SCUNet \\
    \hline
         \multirow{3}{*}{\textbf{CBSD68}} & $\sigma=15$ & 39.83 / 2.06 & - & 39.90 / 1.80 & 40.00 / 1.81 & 39.91 / 1.72 & 39.92 / 1.73 \\
    \cline{2-8}
         & $\sigma=25$ & 39.27 / 2.47 & - & 39.64 / 2.41 & 40.00 / 2.69 & 39.65 / 2.33 & 39.69 / 2.36 \\
    \cline{2-8}
         & $\sigma=50$ & 37.11 / 1.77 & 33.88 / 2.17 & 38.16 / 1.96 & 39.96 / 3.61 & 38.13 / 1.88 & 38.24 / 1.92 \\
    \hline
         \multirow{3}{*}{\textbf{McMaster}} & $\sigma=15$ & 38.84 / 1.22 & - & 39.66 / 0.69 & 40.00 / 0.89 & 39.68 / 0.58 & 39.69 / 0.59 \\
    \cline{2-8}
         & $\sigma=25$ & 38.23 / 1.43 & - & 39.20 / 1.09 & 40.00 / 1.76 & 39.24 / 0.97 & 39.28 / 1.00 \\
    \cline{2-8}
         & $\sigma=50$ & 36.37 / 1.22 & 34.87 / 1.70 & 37.94 / 1.21 & 39.95 / 3.08 & 37.97 / 1.06 & 38.04 / 1.12 \\
    \hline
    \end{tabular}
    }
    \caption{RACDI test results, \textbf{Gaussian Noise} on CBSD68 and McMaster: For each cell of data in this table, the number on the left represents RACDI\_PSNR and the number on the right represents the mean error of RACDI\_PSNR relative to RGCDI\_PSNR. Maximum Mean Error 3.61dB}
    \label{table_noise_cdi}
\end{table*}

\noindent{Calculating $F(F(x, y), y)$}:

\begin{equation}\label{formula:supplementary_property_1_3} 
\begin{aligned}
    &{\mu'}_{A}F(x,y) = F(x,y)\frac{COV(y, F(x,y))}{COV(F(x,y), F(x,y))} \\
    &{\mu'}_{A}F(x,y) = x * COV(y, x)/COV(x, x) \\
    &{\mu'}_{A}F(x,y) = {\mu}_{A}x \\
    &{n'}_{D} = y - {\mu'}_{A}F(x,y) = y - {\mu}_{A}x = {n}_{D}\\
    &{\sigma'}_{D}^{2} = {\sigma}_{D}^{2} \\
    &F( F( x, y ), y ) = \frac{{\mu'}_{A}F(x,y)}{\sqrt{1 + {\sigma'}_{D}^{2} / {\sigma}_{H}^{2}}} \\
    &F( F( x, y ), y ) = \frac{{\mu}_{A}x}{\sqrt{1 + {\sigma}_{D}^{2} / {\sigma}_{H}^{2}}} \\
    &F( F( x, y ), y ) = F( x, y )
\end{aligned}
\end{equation}

% \vspace{5mm}

\noindent \textbf{Cascade Degradation Property.} The $F$ function of cascading multiple degradations is equivalent to cascading multiple $F$ functions of single degradations.

\noindent{Given}:

\begin{equation}\label{formula:supplementary_property_2_1} 
\begin{aligned}
    &y_{1} = D_{1}(x) \\
    &{\mu}_{A1} = COV \left( y_{1}, x \right) / COV \left ( x,x \right ) \\
    &\sigma_{D1}^{2} = COV \left (y_{1} , y_{1} \right ) - {\mu}_{A1} COV \left (y_{1} , x \right ) \\
    &{\mu}_{D1} = 1/\sqrt{1 + {\sigma}_{D1}^{2} / {\sigma}_{H}^{2}} \\
    &F(x, {y}_{1}) = {\mu}_{A1}{\mu}_{D1}x
\end{aligned}
\end{equation}

\begin{equation}\label{formula:supplementary_property_2_1_2} 
\begin{aligned}
    &y_{2} = D_{2}(D_{1}(x)) \\
    &{\mu}_{A2} = COV \left( y_{2}, x \right) / COV \left ( x,x \right ) \\
    &\sigma_{D2}^{2} = COV \left (y_{2} , y_{2} \right ) - {\mu}_{A2} COV \left (y_{2} , x \right ) \\
    &{\mu}_{D2} = 1/\sqrt{1 + {\sigma}_{D2}^{2} / {\sigma}_{H}^{2}} \\
    &F(x, {y}_{2}) = {\mu}_{A2}{\mu}_{D2}x
\end{aligned}
\end{equation}

\noindent{Calculating $F( F(x,y_{1}), y_{2}) $}:

\begin{equation}\label{formula:supplementary_property_2_2} 
\begin{aligned}
    &{\mu'}_{A}F(x,y_{1}) = \frac{F(x,y_{1})COV(y_{2}, F(x,y_{1})))}{COV(F(x,y_{1}), F(x,y_{1}))} \\
    &{\mu'}_{A}F(x,y_{1}) = \frac{{\mu}_{A1}{\mu}_{D1}x * COV(y_{2}, {\mu}_{A1}{\mu}_{D1}x))}{COV({\mu}_{A1}{\mu}_{D1}x, {\mu}_{A1}{\mu}_{D1}x)} \\
    &{\mu'}_{A}F(x,y_{1}) = x * COV(x, y_{2}) / COV(x, x) \\
    &{\mu'}_{A}F(x,y_{1}) = {\mu}_{A2}x \\
    &{n'}_{D} = y_2 - {\mu}_{A2}x = {n}_{D2}\\
    &{\sigma'}_{D}^{2} = \sigma_{D2}^{2} \\
    &F( F(x,y_{1}), y_{2}) = {\mu'}_{A}{\mu}_{D2}F(x,y_{1}) = {\mu}_{A2}{\mu}_{D2}x \\
    &F( F(x,y_{1}), y_{2}) = F(x, {y}_{2})
\end{aligned}
\end{equation}

% \vspace{5mm}

\begin{table}[h]
    \centering
    \resizebox{0.50\textwidth}{!}{
    \begin{tabular}{>
    {\centering\arraybackslash}p{1.0cm}|>
    {\centering\arraybackslash}p{1.0cm}|>{\centering\arraybackslash}p{2.3cm}|>{\centering\arraybackslash}p{2.3cm}|>{\centering\arraybackslash}p{2.3cm}}
    \hline
         \multicolumn{2}{c|}{} & QGAC & FBCNN & GRL-S\\
    \hline
         \multirow{4}{*}{\textbf{LIVE1}} & QF10 & 38.55 / 5.85 & 39.43 / 6.47 & 39.35 / 5.87\\
    \cline{2-5}
         & QF20 & 39.29 / 4.78 & 39.77 / 4.93 & 39.69 / 4.37\\
    \cline{2-5}
         & QF30 & 39.47 / 4.13 & 39.84 / 4.17 & 39.74 / 3.57\\
    \cline{2-5}
         & QF40 & 39.58 / 3.80 & 39.88 / 3.77 & 39.79 / 3.18\\
    \hline
    \end{tabular}
    }
    \caption{RACDI test results, \textbf{JPEG Compression} on LIVE1: For each cell of data in this table, the number on the left represents RACDI\_PSNR and the number on the right represents the mean error of RACDI\_PSNR relative to RGCDI\_PSNR. Maximum Mean Error 6.47dB}
    \label{table_jpeg_cdi}
\end{table}

\noindent \textbf{$\mathbf{RGCDI\_PSNR \ge PSNR}$ (Conditional).} In general, the attenuation coefficients satisfy $0 \le \frac{{\mu}_{A}}{\sqrt{1 + {\sigma}_{D}^{2} / {\sigma}_{H}^{2}}} \le 1 $. Under this condition, $RGCDI\_PSNR$ is bigger than $PSNR$.

\noindent{Given}:

\begin{equation}\label{formula:supplementary_property_3_1} 
\begin{aligned}
    &x = DWT\left[{I}_{x}\right] \\
    &y = DWT\left[{I}_{y}\right] \\
    &\hat{x} = DWT\left[{I}_{\hat{x}}\right] \\
    &{\mu}_{A} = COV \left( y, x \right) / COV \left ( x,x \right ) \\
    &\sigma_{D}^{2} = COV \left (y , y \right ) - {\mu}_{A} COV \left (y , x \right ) \\
    &\mu_{D} = 1/\sqrt{1 + {\sigma}_{D}^{2} / {\sigma}_{H}^{2}} \\
    &\mu_{M} = \arg \min_{\mu_{M}} \left \| \mu_{M}\hat{x} - \mu_{A}\mu_{D}x \right\|
\end{aligned}
\end{equation}

\noindent{Calculating $\| IDWT( \mu_{M}\hat{x}) - IDWT(\mu_{A}\mu_{D}x) \|$}: (Note: Wavelet transform adopts an orthogonal wavelet basis, which has the distance invariance property.)

\begin{equation}\label{formula:supplementary_property_3_2} 
\begin{aligned}
    \| IDWT( \mu_{M}\hat{x}) - IDWT(\mu_{A}\mu_{D}x) \| &= \| \mu_{M}\hat{x} - \mu_{A}\mu_{D}x \| \\
    &\leq \| \mu_{A}\mu_{D}t - \mu_{A}\mu_{D}x \| \\
    &= \mu_{A}^{2}\mu_{D}^{2} \| \hat{x} - x \|
\end{aligned}
\end{equation}

Assume: $\mu_{A}\mu_{D} \leq 1$

\begin{equation}\label{formula:supplementary_property_3_4} 
\begin{aligned}
    &\| IDWT( \mu_{M}\hat{x}) - IDWT(\mu_{A}\mu_{D}x) \| \leq \| \hat{x} - x \| \\
    &\| IDWT( \hat{x}) - IDWT(x) \| = \| \hat{x} - x \| \\
\end{aligned}
\end{equation}

\begin{equation}\label{formula:supplementary_property_3_5} 
\begin{aligned}
    &RGCDI\_MSE \le MSE \\
    &RGCDI\_PSNR \ge PSNR
\end{aligned}
\end{equation}

\section{Attenuation Net Backbones Selection (Detailed Experimental Data)}
Detailed experimental data of attenuation net backbones selection experiments are as follows: Table.\ref{table_ClassicSR_psnr},\ref{table_Noise_psnr},\ref{table_MotionBlur_psnr},\ref{table_JPEG_psnr},\ref{table_Anisotropic_psnr},\ref{table_BlindRestoration_psnr}

\section{RACDI Testing (Supplementary Experimental Results)}

We apply SCUNet as the Attenuation Net backbone, and test the RACDI on more Datsets. Table.\ref{table_classicsr_cdi} shows RACDI test for Classic SR on the BSD100 and Manga109 datasets. Similar to the results on the Urban100 dataset, the maximum mean error is very small, only reaching 0.20dB. Table.\ref{table_noise_cdi} shows RACDI test for Gaussian De-noise on the CBSD68 and McMaster datasets. The maximum mean error is moderate, reaching 3.61dB. Table.\ref{table_jpeg_cdi} shows RACDI test for JPEG Restoration on the LIVE1 dataset. The maximum mean error is larger, reaching 6.47dB.

\end{document}